\documentclass[a4paper,11pt]{article}
\pdfoutput=1

\usepackage{jheppub}

\usepackage[utf8]{inputenc}
\usepackage[T1]{fontenc}
\usepackage{lmodern, amsmath, amssymb, slashed, graphicx, comment,adjustbox,bm}
\usepackage[dvipsnames]{xcolor}
\usepackage{soul}
\usepackage{url}
\usepackage{multirow}
\usepackage{array}
\usepackage{booktabs}
\usepackage{floatrow}
\newfloatcommand{capbtabbox}{table}[][\FBwidth]
\usepackage{braket}
\usepackage{slashed}
\usepackage{mathrsfs}

\usepackage{titlesec}
\titleformat{\subsubsection}[runin]{\normalfont\bfseries}{\thesubsubsection}{1em}{}

\def\figureautorefname~#1\null{Fig.\,#1\null}

\def\equationautorefname~#1\null{Eq.\,(#1)\null}

\newcommand{\La}{\mathcal{L}}

\newcommand{\bpm}{\begin{pmatrix}}
\newcommand{\epm}{\end{pmatrix}}

\newcommand{\apap}{\gamma\phi \to \gamma\phi}
\newcommand{\aaaa}{\gamma\gamma \to \gamma\gamma}

\newcommand{\beq}{\begin{equation} }
\newcommand{\eeq}{\end{equation} }

\title{Positivity bounds in scalar-QED EFT at one-loop level}

\author[a]{Yunxiao Ye,}
\author[a]{Xiao Cao,}
\author[a,c]{Yu-Hang Wu,}
\author[a,b]{Jiayin Gu}

\affiliation[a]{Department of Physics and Center for Field Theory and Particle Physics, \\ Fudan University, Shanghai 200438, China}
\affiliation[b]{Key Laboratory of Nuclear Physics and Ion-beam Application (MOE), \\ Fudan University, 
Shanghai 200433, China}
\affiliation[c]{Department of Physics and Enrico Fermi Institute, University of Chicago, Chicago, IL 60637}

\emailAdd{yxye22@m.fudan.edu.cn}
\emailAdd{xcao23@m.fudan.edu.cn}
\emailAdd{yuhangwu22@m.fudan.edu.cn}
\emailAdd{jiayin\_gu@fudan.edu.cn}

\abstract{
Understanding the implication of positivity bounds on loop-generated dim-8 operator coefficients is a nontrivial task, as these bounds only strictly hold when all the contributions are included in the dispersion relation up to a certain loop order in the UV theory.  
As a step towards more realistic gauge theories such as the Standard Model, in this paper we study the positivity bounds in the Scalar QED Effective Field Theory (EFT) from the scalar-photon scattering ($\apap$) and the photon-photon scattering ($\aaaa$), derived from the dispersion relation of the full one-loop EFT amplitudes.  Assuming the UV theory is weakly coupled and all heavy particles have spin $\leq1$, the leading dim-8 interaction for both amplitudes are generated at the one-loop level in the UV theory.
Gauge invariance imposes strong constraints on the loop structures, while potential IR divergences also require careful treatments.  Our findings reveal that, for $\apap$, while the tree-level bound does not necessarily hold, the one-loop $\beta$-function of the corresponding coefficient always tends to restore the tree-level bound in the IR, unless its actual loop order in the UV theory is further suppressed.  
For $\aaaa$, on the other hand, the tree-level positivity bound is still robust at the one-loop level in the UV theory.  These findings are verified in two example UV models with a heavy scalar extension. 
Importantly, the bounds on the $\beta$-functions that we obtain should be considered as an accidental feature at one loop, rather than a fundamental property of the theory.  
}

\begin{document} 
\maketitle
\flushbottom

\section{Introduction}
\label{sec:Introduction}

Effective Field Theory (EFT) is a powerful framework that connects physics at different scales.
In an EFT, the effects of the underlying ultraviolet (UV) physics can be parameterized as a series of higher dimensional operators in the low-energy effective Lagrangian, suppressed by the cutoff scale $\Lambda$.
A particular important application is the Standard Model Effective Field Theory (SMEFT), given by
\beq
\La_{\rm SMEFT} = \La_{\rm SM} + \sum_{d>4} \frac{c^{(d)}_i}{\Lambda^{d-4}} \mathcal{O}_i^{(d)} \,, \label{eq:Lsmeft}
\eeq
where $\mathcal{O}_i^{(d)}$ is an operator with dimension $d$, and its coefficient $c^{(d)}_i$ is often called the Wilson coefficient.  The SMEFT Lagrangian provides a very useful parameterization to the effects of beyond-Standard-Model (BSM) physics, assuming its scale is significantly higher than the electroweak (EW) scale. 
If the underlying physics is unknown, the Wilson coefficients should be regarded as free parameters to be determined by experiments.  
However, not all Wilson coefficients are free --- 
unitarity, analyticity, and crossing symmetry of the S-matrix impose  
nontrivial constraints, requiring the Wilson coefficients of certain dimension-8 (dim-8) or even higher dimensional operators to be positive~\cite{Adams:2006sv}.  This is commonly denoted as {\it positivity bounds} in an EFT.   
These positivity bounds are of great theoretical interest since they help determine the allowed theory space of the EFT.
In recent years, significant progress has been made in the theoretical study of positivity bounds~\cite{Distler:2006if,Manohar:2008tc,Nicolis:2009qm,Bellazzini:2014waa,Bellazzini:2015cra,Bellazzini:2016xrt,deRham:2017avq,deRham:2017zjm,Bellazzini:2017bkb,Bellazzini:2017fep,deRham:2017xox,deRham:2018qqo,Bellazzini:2019xts,Bellazzini:2019bzh,Wang:2020jxr,Zhang:2020jyn,Alberte:2020jsk,Tokuda:2020mlf,Bellazzini:2020cot,Tolley:2020gtv,Trott:2020ebl,Herrero-Valea:2020wxz,Sinha:2020win,Caron-Huot:2020cmc,Alberte:2020bdz,Arkani-Hamed:2020blm,Li:2021lpe,Caron-Huot:2021rmr,Bern:2021ppb,Aoki:2021ckh,Chiang:2021ziz,Guerrieri:2021tak,Henriksson:2021ymi,Davighi:2021osh,Arkani-Hamed:2021ajd,Du:2021byy,Alberte:2021dnj,Bellazzini:2021oaj,Caron-Huot:2022ugt,Chiang:2022jep,deRham:2022hpx,Chiang:2022ltp,Herrero-Valea:2022lfd,CarrilloGonzalez:2022fwg,Haring:2022sdp,Fernandez:2022kzi,Riembau:2022yse,Hamada:2023cyt,Hong:2023zgm,Bellazzini:2023nqj,CarrilloGonzalez:2023cbf,Xu:2023lpq,Bhat:2023puy,Bertucci:2024qzt,Bittar:2024xuc,Caron-Huot:2024tsk,Beadle:2024hqg,Ye:2024rzr,Bhat:2024agd,Buoninfante:2024ibt,Xu:2024iao,Wan:2024eto,Peng:2025klv,Chang:2025cxc,Beadle:2025cdx,Tokareva:2025rta,Desai:2025alt,Liao:2025npz,Cheung:2025krg,Bhat:2025zex,deRham:2025vaq,Bellazzini:2025shd}. 
The implications of positivity bounds in SMEFT have also been widely studied ~\cite{Low:2009di,Bellazzini:2018paj,Zhang:2018shp,Bi:2019phv, Remmen:2019cyz,Remmen:2020vts,Fuks:2020ujk,Yamashita:2020gtt,Gu:2020ldn,Bonnefoy:2020yee,Remmen:2020uze,Gu:2020thj,Chala:2021wpj, Zhang:2021eeo,Azatov:2021ygj,Li:2022tcz, Li:2022rag, Li:2022aby,Ghosh:2022qqq,Remmen:2022orj,Chen:2023bhu, Gu:2023emi, Davighi:2023acq,  Chala:2023jyx, Chala:2023xjy, Altmannshofer:2023bfk, DasBakshi:2023htx}, which are not only of theoretical interest but also potentially relevant for current and future collider experiments.

Positivity bounds (or at least the simplest version of them) can be obtained by considering the dispersion relations of $2\to 2$ forward elastic amplitudes.  Naturally, their interpretations are most straightforward at the tree level, where the amplitudes can be written as polynomials of Mandelstam variables, and a direct connection can be made between amplitudes and operators, at least in the massless case~\cite{Shadmi:2018xan,Ma:2019gtx,Aoude:2019tzn,Durieux:2019eor,Durieux:2019siw,Gu:2020thj}.
Things are more complicated at the loop level, where logarithmic terms also enters the amplitudes.    
Furthermore, it is well known that even the definitions of renormalized parameters are ambiguous, which depends on the renormalization schemes and renormalization group (RG) running.    
The impacts of loop effects on positivity bounds have been pointed out and investigated in many previous studies~\cite{Bellazzini:2020cot,Herrero-Valea:2020wxz,Arkani-Hamed:2020blm, 
Arkani-Hamed:2021ajd,Chala:2021wpj,Bellazzini:2021oaj,Li:2022aby, Chala:2023jyx,Caron-Huot:2024tsk,Beadle:2024hqg,Ye:2024rzr,Peng:2025klv,Chang:2025cxc,Beadle:2025cdx,Tokareva:2025rta,Desai:2025alt,Liao:2025npz}. 
In particular, Ref.~\cite{Chala:2021wpj} pointed out that loop-generated dim-8 Wilson coefficient could violate the na\"ive tree-level positivity bound.  Following Ref.~\cite{Chala:2021wpj}, Refs.~\cite{Chala:2023jyx,Chala:2023xjy,Liao:2025npz} also further studied the implication of positivity bounds on RG running. 
Taking a different perspective, the recent work Ref.~\cite{Ye:2024rzr} pointed out that it is necessary to include all contributions up to a given loop order in the UV theory for positivity bounds to strictly hold.  
In particular, using scalar EFTs as examples, Ref.~\cite{Ye:2024rzr} demonstrated that certain one loop-contributions to a 4-point forward elastic amplitude correspond to the interference term of the total cross section via the optical theorem.  While the total cross section is positive (which gives the positivity bound), the interference term can obviously take either sign, which means that the corresponding one-loop generated dim-8 Wilson coefficients as well as its $\beta$-function are not subject to any bounds.
Instead, a rigorous positivity bound was derived by considering all contributions to the dispersion relation up to the dim-8 and one-loop levels, in the EFT and loop expansions, respectively. These include the tree-level contribution from the (possibly one-loop generated) dim-8 Wilson coefficients, the one-loop contributions at the dim-8 level which include both log and finite terms, as well as the one-loop contributions at the dim-4 and dim-6 levels which have important effects due to the log terms.  Importantly, the approach in Ref.~\cite{Ye:2024rzr} provides a framework to rigorously derive and interpret the positivity bounds, where the apparent violation of tree-level bounds could be easily understood.        
The implication of positivity bounds at the loop level are also of particular relevance for their experimental probes.  For instance, a observation of the apparent violation of positivity bounds could merely be an indication that the tree-level assumption is false, rather than the break down of the fundamental principles of quantum field theory.

While Ref.~\cite{Ye:2024rzr} focused on scalar theories, it is straightforward to generalize its framework to also include spinning particles.  Nevertheless, when the external particles include gauge bosons (spin-1 particles), the situation can be very different from the scalar case and requires careful studies, since gauge invariance puts stringent restrictions on the possible forms of couplings and amplitude structures.  
In particular, the gauge interactions come from kinetic terms, so they could not mix particles with different masses.\footnote{For instance, assuming kinetic terms are canonical, there is no 3-point interaction of the form $\gamma \phi \Phi$, where $\gamma$ is a gauge boson (photon), $\phi$ is a light scalar and $\Phi$ is a heavy scalar.  The same also applies to other particles.} For the amplitude $\gamma x \to \gamma x$ where $x$ denotes a light particle, one could immediately conclude that it cannot have a tree-level $s$-channel exchange of a heavy particle.   
As a first step towards a more complete understanding of positivity bounds for gauge theories and the SM, in this paper we consider the simplest gauge theory --- scalar QED and its EFT, and study the positivity bounds in this theory at the one-loop level.  
A number of important assumptions are made in our study.   
First of all, we always work in 3+1 spacetime dimensions.
We assume that the UV theory is a weakly coupled quantum field theory (QFT), so that the classification of loop orders is meaningful to begin with.  We further assume all the particles in the UV theory -- massless or massive -- have spin less or equal to one. 
Particles with higher spins introduce additional kinematic dependence in their couplings.
In many cases, their interactions are non-renormalizable which means the theory needs further UV completion.  Note this also means that we do not consider any gravitational effects. In fact, the impacts of the $t$-channel graviton exchange on positivity bounds is a highly nontrivial topic and has been extensively studied in the past~\cite{Alberte:2020jsk,Tokuda:2020mlf,Herrero-Valea:2020wxz,Alberte:2020bdz}.
Importantly, under these assumptions, the $\gamma \phi  \to \gamma \phi $ and $\gamma \gamma \to \gamma \gamma$ amplitudes we consider can only be generated at the one-loop level in the UV theory. 
Note that, these assumptions do introduce some model dependence in our results. 
In particular, they exclude UV models that are generated by strong dynamics (QCD-like theory), or contain compactified extra dimension, both of which predict massive higher-spin particles. One major obstacle for applying our framework to such theories is that the matching between EFT and the UV physics is not fully understood -- strictly speaking, it would require solving non-perturbative gauge theories or quantum gravity.  However, we note that such UV theories typically predict a massive spin-2 particle which generates a dominant tree-level contribution to the processes we study (see {\it e.g.} discussions in Ref.~\cite{Bellazzini:2018paj}), in which case the tree-level bound is expected to hold.  As such, the assumptions we make characterize a rather generic situation where it is important to consider positivity bounds beyond the tree level. 
We also assume the only light degrees of freedom are the massless photon and a complex scalar.  For simplicity (in particular for the loop calculations), we assume the scalar is also massless, except when the dispersion relation contain infrared (IR) divergences of the form $\log(-t)$ in the forward limit, in which case a small mass is introduced as a regulator.   

We will focus on the elastic processes $\gamma \phi  \to \gamma \phi $ and $\gamma \gamma \to \gamma \gamma$, where the initial and final state photons ($\gamma$) have the same polarization, while two incoming photons can have different polarizations.  
Under the assumptions above, these two amplitudes could only be generated at the one-loop level in the UV theory ({\it i.e.}, they cannot be generated by a tree-level heavy particle exchange), which makes the loop-level positivity bounds particularly relevant.\footnote{  
On the other hand, the $\phi\phi\to\phi\phi$ amplitude can be generated at the tree level.  Scalar amplitudes have also been studied in Ref.~\cite{Ye:2024rzr}, though in a different EFT. As such, we decide to focus on the amplitudes involving photons.}   
They are also free of simple poles of the form $1/t$ which can be problematic in the forward limit.   
In each process, we compute the relevant dispersion relation up to the dim-8 and one-loop levels. 
Due to the form of gauge interactions, the results exhibit a number of important features different from the scalar EFT case in Ref.~\cite{Ye:2024rzr}. First, their dispersion relations 
generally exhibit IR divergences in the form of $1/\epsilon$ from dimensional regularization, or $\log(-t)$ in the forward limit, both of which require careful treatments.  
Second, for the two processes considered here, one could subtract the dim-4 contributions from the dispersion relation, which still remains  positive.
Interestingly, for $\gamma \phi  \to \gamma \phi$, we find that the $\beta$-function of the corresponding dim-8 coefficient ($c_{F^{2}D^{2}\phi^{2}}^{(1)}$ in \autoref{eq:La8}) is indeed subject to a positivity bound, at least at the one-loop level. 
The dim-8 coefficient $c_{F^{2}D^{2}\phi^{2}}^{(1)}$ itself does not necessarily obey the tree-level positivity bound, but the $\beta$-function always tend to restore the tree-level bound of $c_{F^{2}D^{2}\phi^{2}}^{(1)}$ in the IR.  
These observations are further checked and verified in two simple UV models, which provide useful insights.  
For $\gamma \gamma \to \gamma \gamma$, on the other hand, all Wilson coefficients that contribute at the one-loop level are themselves generated at the one-loop level in the UV, so the one-loop contributions in the EFT are actually two-loop effects in the UV theory. In this case, the tree-level positivity bounds manifestly hold at the one-loop level in the UV theory.  Nevertheless, we will work out the full one-loop dispersion relations in the EFT and discuss the issues that arise when trying to apply them beyond the one-loop level in the UV theory.

The rest of this paper is organized as follows. In \autoref{sec:pos}, we review the derivation of the dispersion relation that leads to the positivity bounds. 
In \autoref{sec:sQED}, we lay down the theory framework of the scalar-QED EFT and also demonstrate the absence of $1/t$ poles in the  amplitudes we consider. In \autoref{sec:phosca}, we calculate the full one-loop $\gamma \phi \to \gamma \phi$ amplitude and derive the corresponding positivity bounds, which are verified in two different UV models with a heavy scalar extension.    
In \autoref{sec:phopho} we perform the same analysis on the $\aaaa$ amplitude.  
The conclusion is drawn in \autoref{sec:con}. 
Additional details of our calculations are provided in the Appendix, where \autoref{sec:masint} shows the explicit forms of the one-loop scalar integrals, \autoref{sec:matching} contains the full one-loop matching results, and the results for the dim-4 one-loop 
$\aaaa$ amplitudes are shown in \autoref{sec:aaaadim4}.


\section{The dispersion relation}
\label{sec:pos}

In this section, we will briefly review the derivation of the dispersion relation which leads to the positivity bounds that we are interested in. Our starting point is the elastic scattering of two particles which we denote as $a$ and $b$.  To simplify loop calculations, we consider the limit where both particle $a$ and $b$ are massless (or at least the mass is negligible compared with other finite scales).  In general, its amplitude is a function of the Lorentz invariant Mandelstam variables $s$, $t$ and $u$.  
The forward elastic amplitude is obtained by taking the $t\to0$ limit, 
\begin{equation}
    A_{ab \to ab}(s) = \mathcal{A}(ab \to ab) |_{t \to 0} \,,
\end{equation}
which can be considered as a function of $s$ due to the relation $s+t+u =0$.
For the dispersion relation, we use the version proposed in Ref.~\cite{Herrero-Valea:2020wxz} (and also adopted in Refs.~\cite{Chala:2023jyx,Chala:2023xjy,Ye:2024rzr}) which has several advantages when applying to loop-level amplitudes.  
This version specifically requires the amplitude to be invariant under the $s\leftrightarrow u$ exchange.  
For spinning particles, the forward elastic amplitude under the $s\leftrightarrow u$ exchange transforms as (see {\it e.g.} Ref.~\cite{Bellazzini:2016xrt}) 
\begin{equation}
    A(a^{h_{1}}_{\rho_{1}} b^{h_{2}}_{\rho_{2}} \to a^{h_{1}}_{\rho_{1}} b^{h_{2}}_{\rho_{2}}, s) \to A( \bar{a}^{-h_{1}}_{\bar{\rho}_{1}} b^{h_{2}}_{\rho_{2}} \to \bar{a}^{-h_{1}}_{\bar{\rho}_{1}} b^{h_{2}}_{\rho_{2}}, u=-s ) \,,
\end{equation}
where the incoming and outgoing particle $a$ are swapped, $\bar{a}$ is the anti-particle of $a$, $h_{i}$ denotes the helicity, and $\rho_{i}$ denote the collections of internal quantum numbers.  Note the crossing changes $a$ to $\bar{a}$ and also flips its helicity and internal quantum numbers (if any). 
The $s\leftrightarrow u$ crossing symmetry is a nontrivial requirement which can be satisfied by the $\apap$ and $\aaaa$ amplitudes we consider.

\begin{figure}[tbp]
	\centering
	\includegraphics[width=0.7\textwidth]{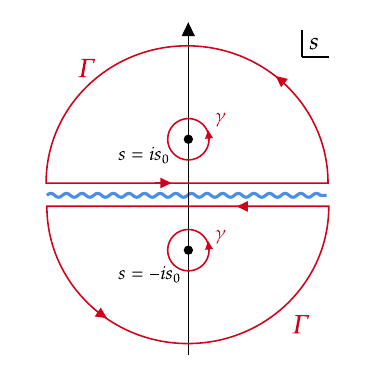}
	\caption{Contours in the complex $s$-plane. The blue line represents the branch cut on the real axis.
    The contours $\gamma$ around $s=\pm is_0$ are deformed to the contours $\Gamma$, where the big semi-circles of $\Gamma$ is at $|s| \to \infty$.} 
    \label{fig:contour}
\end{figure}

We consider the following quantity 
\begin{equation}
\label{eq:Sigma1}
    \Sigma \equiv \oint_{\gamma} \frac{ds}{ 2\pi i } \frac{ s^{3} A(s) }{ (s^{2} + s_{0}^{2})^{3} } = \Bigg(\ \oint\limits_{ s = i s_{0} } + \oint\limits_{ s = - i s_{0} } \Bigg) \frac{ds}{ 2\pi i } \frac{ s^{3} A(s) }{ (s^{2} + s_{0}^{2})^{3} }\,,
\end{equation}
where $\gamma$ indicates the sum of the two contours around $s=i s_0$ and $s= - i s_0$, with $s_0$ assumed to be real and positive. 
Note the two contour integrals also equal as a result of the $s\leftrightarrow u$ crossing symmetry.  We then deform the contours from $\gamma$ to $\Gamma$ where the two semi-circles extends to infinity, as shown in \autoref{fig:contour}. 
$A(s)$ is analytic in the entire complex $s$-plane except for poles and branch cuts on the real axis.  For massless particles, the branch cut extends along the entire real axis.  It is worth mentioning that the original positivity bounds~\cite{Adams:2006sv} require an ``analytic gap'' on the real axis to deform the contour to the entire $s$-plane.  This problem is avoided here by considering two separate contours on the upper and lower plane.  We also assume the contributions from the two semi-circles at infinity vanishes.\footnote{This condition can be regarded as a generalization of the Froissart-Martin bound~\cite{Froissart:1961ux,Martin:1962rt}, which requires $A(s) < s \log^2(s)$ when $s\to \infty$, assuming the theory has a mass gap. While scalar QED contains massless particles, we still expect this bound to hold in generic UV completions without gravitational effects.}
Therefore, the $\Sigma$ in \autoref{eq:Sigma1} can be rewritten as the difference between the line integral above and below the real axis: 
\begin{equation}
    \begin{split}
    \label{eq:Sigma2}
        \Sigma =\int_{-\infty}^{\infty} \frac{ds}{2\pi i} \frac{ s^{3} \left( A(s+i\varepsilon) - A(s-i\varepsilon) \right) }{ \left(s^{2} + s_{0}^{2}\right)^{3} } \,, 
    \end{split}
\end{equation}
where $\varepsilon$ is taken to be infinitesimal.
By applying the the crossing symmetry $A(s) = A(-s)$ and Schwarz principle $A^{\ast}(s) = A(s^{\ast})$, we can write
\begin{equation}
    \begin{split}
        \Sigma = \int_{0}^{\infty} \frac{ds}{\pi i} \frac{ s^{3} \left( A(s+i\varepsilon) - A^{\ast}(s+i\varepsilon) \right) }{ \left(s^{2} + s_{0}^{2}\right)^{3} } = \frac{2}{\pi} \int_{0}^{\infty} ds\, \frac{s^{3}\, {\mathrm{Im}} \big[A(s)\big] }{ \left(s^{2} + s_{0}^{2}\right)^{3} } \,. 
    \end{split}
\end{equation}  
Finally, by applying the optical theorem $\text{Im}[A(s)]=s\sigma(s)$ where $\sigma(s)$ is the total cross section for the scattering of $a$ and $b$, we obtain
\begin{equation}\label{eq:Sigmapos}
    \Sigma \equiv \oint_{\gamma} \frac{ds}{ 2\pi i } \frac{ s^{3} A(s) }{ (s^{2} + s_{0}^{2})^{3} } =  \frac{2}{\pi} \int_{0}^{\infty} ds\, \frac{ s^{4} \sigma(s) }{ \left(s^{2} + s_{0}^{2}\right)^{3} } \geq 0 \,.
\end{equation}

At the one-loop level, the massless forward amplitude $A(s)$ also contain terms proportional to $\log(s)$ in addition to polynomial terms of $s$. 
Some useful formulae are given here:
\begin{equation}
\label{eq:ctfor}
    \begin{split}
        \oint_{\gamma} \frac{ds}{ 2\pi i } \frac{ s^{3} }{ (s^{2} + s_{0}^{2})^{3} }  = 0 \, , \hspace{2.2cm}
        \oint_{\gamma} \frac{ds}{ 2\pi i } \frac{ s^{3} }{ (s^{2} + s_{0}^{2})^{3} } \log(s) &= - \frac{1}{4} \frac{1}{s_{0}^{2}} \, , \\
        \oint_{\gamma} \frac{ds}{ 2\pi i } \frac{ s^{3} }{ (s^{2} + s_{0}^{2})^{3} } s  = 0 \, , \hspace{2cm}
        \oint_{\gamma} \frac{ds}{ 2\pi i } \frac{ s^{3} }{ (s^{2} + s_{0}^{2})^{3} } s \log(s) &= \frac{ 3 \pi }{ 16 s_{0} } \, , \\
        \oint_{\gamma} \frac{ds}{ 2\pi i } \frac{ s^{3} }{ (s^{2} + s_{0}^{2})^{3} } s^2  =  1 \, , \hspace{1.7cm}
        \oint_{\gamma} \frac{ds}{ 2\pi i } \frac{ s^{3} }{ (s^{2} + s_{0}^{2})^{3} } s^2 \log(s) &= \frac{3}{4} + \log(s_{0}) \,.
    \end{split}
\end{equation}  
In particular, for polynomials of $s$, only the $s^2$ (or even higher powers of $s$) term in the amplitude contributes to $\Sigma$, but with an extra factor of $\log(s)$ all terms could contribute.  This means that the dim-4 and dim-6 one-loop contributions generally enters the dispersion relation, which has important implications as pointed out in Ref.~\cite{Ye:2024rzr}.    

It is also important to note that, while the dispersion relation in \autoref{eq:Sigmapos} is valid for any $s_0$, in the EFT calculation we always truncate at the dim-8 level. Dim-10 and even higher dimensional operators can contribute to \autoref{eq:Sigmapos}, but are suppressed by additional powers of $s_0/\Lambda^2$.  To ensure the validity of the EFT expansion, we require that $s_0 \ll \Lambda^2$.


\section{The Scalar-QED EFT framework}
\label{sec:sQED}

We consider the scalar-QED EFT which includes a $U(1)$ gauge boson $A_{\mu}$ referred to hereafter as the ``photon'', and a complex scalar field $\phi$. For simplicity of the loop calculation, we consider $\phi$ to be massless, and its mass term $m^2$ only appears later as a IR regulator.   
We consider the Lagrangian up to the dim-8 level, given by
\beq
\La = \La_{[\mathcal{O}]\leq 4} + \La_{6} + \La_{8} \,.
\eeq
Note that there is no dim-5 or dim-7 operator in this theory.  The renormalizable part of the Lagrangian is
\begin{equation}
\label{eq: scalar qed d<=4}
\mathcal{L}_{[\mathcal{O}]\leq 4} = - \frac{1}{4} F^{\mu \nu} F_{\mu \nu}+( D^{\mu} \phi )^{\dagger} ( D_{\mu} \phi )- \frac{1}{4} \lambda_{1} ( \phi^{\dagger} \phi )^{2} \, ,
\end{equation}
where $D_{\mu} \phi = \partial_{\mu} \phi - i e q_{\phi} A_{\mu} \phi$ and $q_{\phi}$ is the charge of $\phi$ which we simply set to 1 without loss of generality.  We also label the scalar quartic coupling $\lambda_1$ to distinguish it from additional scalar couplings in the UV models in \autoref{subsec:topdow}. 
The dim-6 Lagrangian is 
\begin{equation}
    \begin{split}
        \mathcal{L}_{6} \cdot \Lambda^2 = c_{\phi^{6}} (\phi^{\dagger}\phi)^3 + c_{D^{2}\phi^{4}} \phi^{\dagger}\phi(D_{\mu}\phi)^{\dagger}D^{\mu}\phi + c_{F^{2}\phi^{2}} \phi^{\dagger}\phi F^{\mu \nu} F_{\mu \nu} \, ,
    \end{split} \label{eq:La6}
\end{equation}
and the dim-8 one is 
\begin{equation}
    \begin{split}
        \mathcal{L}_{8} \cdot \Lambda^4 =& c_{F^{4}}^{(1)} \, (F_{\mu\nu}F^{\mu\nu})^2 + c_{F^{4}}^{(2)} \, ( F_{\mu \nu} \tilde{F}^{\mu \nu} )^2 \\
        &+ c_{F^{2}D^{2}\phi^{2}}^{(1)} \, (D_{\mu}\phi)^{\dagger}(D_{\nu}\phi)F^{\mu\rho}F^{\nu}_{\rho} + c_{F^{2}D^{2}\phi^{2}}^{(2)} \, ( D^{\mu} \phi )^{\dagger} ( D_{\mu} \phi )F_{\nu\rho}F^{\nu\rho} \\
        &+ c_{D^{4}\phi^{4}}^{(1)} \, ((D_{\mu}\phi)^{\dagger})^2 (D_{\nu}\phi)^2 + c_{D^{4}\phi^{4}}^{(2)} \, \phi^{\dagger}\phi(D_{\mu}D_{\nu}\phi)^{\dagger}(D_{\mu}D_{\nu}\phi) \\
        &+ c_{F^{2}\phi^{4}} \, (\phi^{\dagger}\phi)^2 F_{\mu\nu}F^{\mu\nu} + c_{D^{2}\phi^{6}} \, (\phi^{\dagger}\phi)^2(D_{\mu}\phi)^{\dagger}D^{\mu}\phi + c_{\phi^{8}} \, (\phi^{\dagger}\phi)^4 \, ,
    \end{split} \label{eq:La8}
\end{equation}
where $\tilde{F}_{\mu \nu} = \epsilon_{\mu \nu \rho \sigma} F^{\rho \sigma}/2$.  
Both $\La_6$ and $\La_8$ contain a complete set of non-redundant CP-even operators.   We will compute the $\apap$ and $\aaaa$ amplitudes up to one loop and $\mathcal{O}(1/\Lambda^4)$ order.  The coefficients $c_{\phi^6}$, $c_{\phi^8}$, $c_{D^{2}\phi^{6}}$ and $c_{F^{2}\phi^{4}}$ do no contribute to the $\apap$ and $\aaaa$ amplitudes at this level.\footnote{For $c_{F^{2}\phi^{4}}$, one could connect two scalars  to form a diagram for the $\apap$ amplitude.  However, for massless $\phi$ this is a scaleless loop and vanishes in dimensional regularization.}  While $c_{D^{2}\phi^{4}}$ could contribute to $\apap$ at the one-loop level, its contribution is proportional to $t$ which vanishes in the forward limit.  Furthermore, $c_{D^{2}\phi^{4}}$ does not contribute to the helicity amplitude $\gamma^+ \phi \to \gamma^+ \phi$ that we consider.
For the rest dim-6 and dim-8 operator coefficients, 
\beq
c_{F^{2}\phi^{2}} \,, \quad
c_{F^{4}}^{(1)} \,, \quad
c_{F^{4}}^{(2)} \,, \quad
c_{F^{2}D^{2}\phi^{2}}^{(1)}  \,, \quad
c_{F^{2}D^{2}\phi^{2}}^{(2)} \,,  \quad 
c_{D^{4}\phi^{4}}^{(1)} \,, \quad
c_{D^{4}\phi^{4}}^{(2)} \,, \label{eq:c1loop}
\eeq
only the dim-8 4-scalar coefficients $c_{D^{4}\phi^{4}}^{(1)}$ and $c_{D^{4}\phi^{4}}^{(2)}$ could be generated at the tree level, while the rest are only generated at the one-loop level under the assumption that heavy particles have spin $\leq 1$.  This can be easily deduced since there is no possible tree-level heavy particle exchange to the corresponding $\apap$ and $\aaaa$ amplitudes.

\begin{figure}[t]
	\centering
	\includegraphics[width=0.4\textwidth]{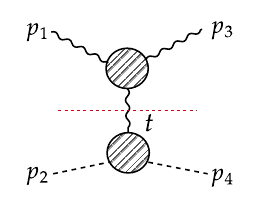}
	\caption{If the $\apap$ amplitude contain a $1/t$ pole, it would correspond to a massless $t$-channel propagator (represented by the solid line).  In the $t\to 0$ limit the massless propagator goes on-shell and the amplitude factorizes into two sub-amplitudes (cut by the red dashed line). The shadowed circle represents a generic vertex that include both tree-level and loop diagrams. }
    \label{fig: t channel}
\end{figure}

One nice feature of the $\apap$ and $\aaaa$ amplitudes we consider is that they are free from any $1/t$ poles, which could potentially make the forward amplitude ill-defined.\footnote{A $1/t$ pole is not necessarily problematic for the (twice-subtracted) dispersion relation we consider.  For instance, a tree-level amplitude $\sim s/t$ simply does not contribute to $\Sigma$ according to \autoref{eq:ctfor}.} 
This statement holds at all loop order due to the following simple arguments.
As illustrated in \autoref{fig: t channel}, the $1/t$ pole can only be produced by a massless particle in the $t$-channel, which in our case would be either the photon or the scalar $\phi$.\footnote{More formally speaking, an amplitude with a $1/t$ pole factorize into two sub-amplitudes in the $t\to 0$ limit, where the intermediate massless particle is taken on-shell.  This is represented by the red dash line that cuts the $t$-channel propagator in \autoref{fig: t channel}.}  
A $t$-channel photon exchange has no contribution as a consequence of \textit{Furry's theorem}, which forces the 3-photon vertex to vanish.  A $t$-channel exchange of $\phi$ is forbidden by the $U(1)$ charge conservation. 
Note that the arguments here rely only on the assumption that $\gamma$ and $\phi$ are the only light degrees of freedom, and remain valid even with the presence of heavy particles with spin $>1$.  
On the other hand, IR divergences of the form $\log t$ will appear in our calculation, which will be discussed in the next section.

We work in the $\overline{\rm MS}$ scheme and employ the Passarino-Veltman reduction~\cite{Passarino:1978jh} to calculate the one-loop amplitudes.
It is well known that any one-loop integral can be reduced to a linear combination of a basis of scalar integrals, $B_{0}$, $C_{0}$, $D_{0}$ which correspond to bubble, triangle, and box scalar integrals, plus rational terms $\mathscr{R}$ which arise if the dimensional parameter $D$ is involved in the reduction and are related to UV singularities.
\begin{equation}
    \mathcal{A}^{\text{1-loop}} = \sum_{\text{box}} d_{4} D_{0} + \sum_{\text{triangle}} d_{3} C_{0} + \sum_{\text{bubble}} d_{2} B_{0} + \mathscr{R} \,,
\end{equation}
where the integral coefficients $d_4$, $d_3$, and $d_2$ and rational term $\mathscr{R}$ are rational functions of external kinematic, and the integrated results of master integrals are provided in \autoref{sec:masint}.
In this paper, the calculation of one-loop amplitudes is assisted by the \texttt{FeynRules}~\cite{Christensen:2008py,Alloul:2013bka}, \texttt{FeynArts}~\cite{Hahn:2000kx}, \texttt{FeynCalc}~\cite{Mertig:1990an,Shtabovenko:2016sxi,Shtabovenko:2020gxv,Shtabovenko:2023idz}, \texttt{FeynHelpers}~\cite{Shtabovenko:2016whf}, \texttt{Package-X}~\cite{Patel:2015tea,Patel:2016fam} packages.


\section{Photon-scalar scattering}
\label{sec:phosca}

\subsection{Implication of the dispersion relation}
\label{subsec:Imp}

In this section, we calculate the $\apap$ amplitude and apply the dispersion relation in \autoref{eq:Sigmapos} to it.  
Note again that we assume the initial and final state photons have the same polarization so that the amplitude is elastic.  
We focus on the case where the photon has a positive helicity, with the corresponding amplitude labeled as $\mathcal{A}_{+ \phi \to + \phi}$.  Having the photon with the opposite helicity, or $\phi^\dagger$ instead of $\phi$, does not give independent bounds due to the crossing symmetry of the amplitude.  We also find that having linearly polarized photons would still give the same bounds. 
The dispersion relation in \autoref{eq:Sigmapos} assumes that the amplitude is invariant under the $s \leftrightarrow u$ exchange, which transforms $\mathcal{A}_{+ \phi \to + \phi}$ to $\mathcal{A}_{+ \phi^\dagger \to + \phi^\dagger}$.  This indeed holds and is guaranteed by the charge conjugation symmetry in scalar QED.

The full tree-level amplitude up to the dim-8 level is given by\footnote{While it is simpler to directly use the polarization vector in the forward direction $\epsilon^{+}_{\mu}=(0,1,i,0)$ for the tree-level forward amplitudes, for loop-level amplitudes one needs to first perform the calculation at arbitrary polar angle $\theta$ and azimuthal angle $\varphi$ before taking the forward limit, as this properly accounts for the angular dependence of the kinematics in loop integrals. In our convention, the circular polarizations are chosen as $\epsilon^{\pm}_{\mu}(p) = \frac{1}{\sqrt{2}} ( 0 , \cos \theta \cos \varphi \mp i \sin \varphi , \cos \theta \sin \varphi \pm i \cos \varphi, - \sin \theta )$.}
\begin{equation}
    \mathcal{A}^{\text{tree}}_{+ \phi \to + \phi} = - 2 e^2 + \frac{1}{\Lambda^{4}} \frac{1}{2} c_{F^{2}D^{2}\phi^{2}}^{(1)} su \,.
\end{equation}
While the dim-4 amplitude contain diagrams with an $s$ or $u$ channel scalar propagator, the sum of all diagrams is simply $-2e^2$ for the chosen polarization vectors.
The tree-level helicity amplitude $\mathcal{A}_{+ \phi \to + \phi}$ receives no contribution from the dim-6 coefficient $c_{F^{2}\phi^{2}}$ or the dim-8 coefficient $c_{F^{2}D^{2}\phi^{2}}^{(2)}$, both of which contribute to $\apap$ where the initial and final state photon are in opposite helcities.   
Only the dim-8 coefficient $c_{F^{2}D^{2}\phi^{2}}^{(1)}$ contributes to the dispersion relation, which gives the tree-level positivity bound
\begin{equation}
\label{eq:phi-gamma-treebound}
    c_{F^{2}D^{2}\phi^{2}}^{(1)} \leq 0 \, .
\end{equation}

We now go to the one-loop level.  We will first be agnostic about the loop orders in the UV theory ({\it i.e.} whether an operator is generated at tree level or loop level), and compute the complete one-loop amplitude in the EFT. Starting from the dim-4 contributions, we have
\begin{equation}
\begin{split}
\mathcal{A}^{[4],\text{1-loop}}_{+ \phi \to + \phi}=&
-\frac{e^4}{8 \pi^2}\frac{1}{s u} \Big( -2 s u (B_0(s) + B_0(u)) + (s^2 + t^2 + u^2) t C_0(t)\\
& \quad -2 t s^2 C_0(s) -2 t u^2 C_0(u)
+ s^2 t^2 D_0(s,t) + u^2 t^2 D_0(t,u) \Big) \,.
\end{split}
\end{equation}
Note that the UV divergences in $B_0$ can be absorbed by counterterms, while the IR divergences in $C_0$ and $D_0$ remain.

Since the dim-6 tree-level amplitude vanishes, the dim-6 one-loop amplitude is finite and contains only a rational term:
\begin{equation}
\mathcal{A}^{[6],\text{1-loop}}_{+ \phi \to + \phi} = \frac{1}{\Lambda^2} \frac{1}{4\pi^2} e^{2} c_{F^2 \phi^2}(s+u)\,,
\end{equation}
which vanishes in the forward limit.  Note also that a rational dim-6 amplitude would not contribute to the dispersion relation even if it is nonzero in the forward limit, as shown in \autoref{eq:ctfor}.  

The dim-8 one-loop amplitude is\footnote{Note also that $c_{F^{2}D^{2}\phi^{2}}^{(2)}$ only contributes to the rational part due to its helicity structure.  See {\it e.g.} the discussion in Ref.~\cite{Craig:2019wmo}.}
\begin{equation}
\begin{split}
\mathcal{A}^{[8],\text{1-loop}}_{+ \phi \to + \phi}=&-\frac{s u}{96\pi^2\Lambda^4} \Big( (6 e^2 c_{F^2 D^2\phi^2}^{(1)} + 8 c_{F^2\phi^2}^2)(B_0(s)+B_0(u)) \\
& \quad +e^2 ( 2 c_{D^4\phi^4}^{(1)} + c_{D^4\phi^4}^{(2)}+ 18 c_{F^2 D^2 \phi^2}^{(1)}+ 96 c_{F^{4}}^{(1)} + 96 c_{F^{4}}^{(2)}) B_0(t)\\
& \quad + 6 e^2 c_{F^2 D^2\phi^2}^{(1)} t C_0(t) \Big) + \mathscr{R} \,,
\end{split}
\end{equation}
where the rational part $\mathscr{R}$ is
\begin{equation}
    \begin{split}
        \mathscr{R} =& -\frac{1}{576\pi^2\Lambda^4}\Bigg(s u \Big(14 e^2 c_{D^4\phi^4}^{(1)}+ 7 e^2 c_{D^4\phi^4}^{(2)} - 1152 e^2 c_{F^{4}}^{(2)} + 16 c^2_{F^2\phi^2}\Big)\\
        & \quad \quad + 6 e^2 c_{F^2 D^2 \phi^2}^{(1)}(t^2 + 2 s u)+ 24 e^2 c_{F^2 D^2\phi^2}^{(2)} (2(s^2+u^2)-t^2) \Bigg)\,.
    \end{split}
\end{equation}
Note that $\mathcal{A}^{[8],\text{1-loop}}_{+ \phi \to + \phi}$ also contain a dim-6-squared contribution (proportional to $c_{F^2\phi^2}^2$) which is also at the $\sim 1/\Lambda^4$ order.   
It is also straightforward to calculate the beta function of $c_{F^2 D^2\phi^2}^{(1)}$, which is\footnote{Note that when computing the beta function through dimensional regularization, one must retain the scaleless integral $B_{0}(0)$ in order to correctly extract the UV divergent part.}
\begin{equation}
\label{eq:beta1}
\begin{split}
\beta(c^{(1)}_{F^2 D^2\phi^2}) &\equiv \mu \frac{d}{d\mu} c^{(1)}_{F^2 D^2\phi^2} \\ &= \frac{1}{24\pi^2}\left(e^2 ( 2 c_{D^4 \phi^4}^{(1)} + c_{D^4\phi^4}^{(2)} + 96 c_{F^{4}}^{(1)} + 96 c_{F^{4}}^{(2)} + 24 c_{F^2 D^2\phi^2}^{(1)})+ 16 c_{F^2\phi^2}^2\right)\,.
\end{split}
\end{equation}

\begin{figure}[t]
	\centering
	\includegraphics[width=0.8\textwidth]{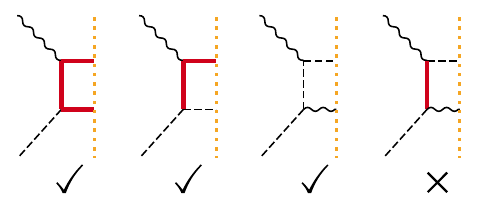}
	\caption{An illustration of some typical contributions to the one-loop forward elastic amplitude of $\gamma^+\phi \to \gamma^+\phi$ in the UV theory.  Only the left halves of the one-loop diagrams are drawn, which correspond to the diagrams for the tree-level $\sigma(\gamma^+ \phi \to {\rm 2~particles} )$.  The thick red line represents a generic heavy particle.  The heavy particle only contributes where at least one of the final state particles is heavy.  $\sigma(\gamma^+ \phi \to \gamma \phi)$ receives no contribution from the heavy particle at the tree level since it is not allowed by the structure of the gauge interaction. Note that we assume the UV theory is weakly coupled and heavy particles have spin $\leq 1$. Not all diagrams are shown here. 
    } 
    \label{fig:apsigma}
\end{figure}

Combining all the contributions and plugging $\mathcal{A}_{+ \phi \to + \phi}$ in the left-hand side (LHS) of \autoref{eq:Sigmapos}, we obtain (in the $t\to 0$ limit)
\begin{equation}
    \begin{split}
    \label{eq:phi-gamma-sigma}
        \Sigma =& \frac{e^4}{8 \pi^2} \frac{ 1 }{ s_{0}^{2} } - \frac{1}{\Lambda^{4}} \frac{1}{2} c_{F^{2}D^{2}\phi^{2}}^{(1)} \\ 
        &+ \frac{ 1 }{ \Lambda^{4} } \frac{1}{96 \pi^2} \Bigg[ 6 e^2 c_{F^2 D^2\phi^2}^{(1)} \left( \frac{1}{\epsilon^2} - \frac{1}{\epsilon} \log \left( \frac{ -t }{ \mu^{2} } \right) + \frac{1}{2} \log^{2} \left( \frac{ -t }{ \mu^{2} } \right) - \frac{1}{2} \zeta_{2} \right) \\
        & \quad \quad + e^2 ( 2 c_{D^4\phi^4}^{(1)} + c_{D^4\phi^4}^{(2)}+ 18 c_{F^2 D^2 \phi^2}^{(1)}+ 96 c_{F^{4}}^{(1)} + 96 c_{F^{4}}^{(2)} ) \left( - \log \left( \frac{ -t }{ \mu^{2} } \right) + 2 \right) \\
        & \quad \quad + (6 e^2 c_{F^2 D^2\phi^2}^{(1)} + 8 c_{F^2\phi^2}^2) \left( - \frac{3}{2} - 2 \log\left( \frac{s_0}{\mu^2} \right) + 4 \right) \Bigg] \\
        &+ \frac{1}{576\pi^2\Lambda^4}\Bigg[ 14 e^2 c_{D^4\phi^4}^{(1)}+7 e^2 c_{D^4\phi^4}^{(2)} - 1152 e^2 c_{F^{4}}^{(2)} + 16 c^2_{F^2\phi^2} + 12 e^2 c_{F^2 D^2 \phi^2}^{(1)} - 96 e^2 c_{F^2 D^2 \phi^2}^{(2)} \Bigg]\,,
    \end{split}
\end{equation}
where $\zeta_2 = \frac{\pi^2}{6}$ and the explicit forms of the master integrals in \autoref{sec:masint} has been substituted in. \autoref{eq:phi-gamma-sigma} is quite complicated and contain serval contributions --- some of which are divergent --- that could potentially prevent us from extracting any useful positivity bounds.  First, there is a dim-4 one-loop contribution $\frac{e^4}{8 \pi^2} \frac{ 1 }{ s_{0}^{2} }$ which dominates at small $s_0$.  On the other hand, the dim-6 contribution vanishes as we just mentioned. 
For the scalar-EFT case in Ref.~\cite{Ye:2024rzr}, the dim-4 and dim-6 contributions play an important role in the interpretation of the positivity bound.  
However, here we can subtract the dim-4 contribution and the remain part of $\Sigma$ is still positive, assuming that the UV theory is weakly coupled and heavy particles have spin $\leq 1$.
To see this, we note that the one-loop forward elastic amplitude on the LHS of the dispersion relation \autoref{eq:Sigmapos} corresponds to $2\to 2$ tree-level cross section on the right hand side (RHS), which can be separated into two parts,
\beq 
\sigma(\gamma^+ \phi \to {\rm 2~particles}) =  \sigma(\gamma^+ \phi \to {\rm 2~light~particles}) + \sigma(\gamma^+ \phi \to {\rm else}) \,, \label{eq:cs2to2}
\eeq
where the second part $\sigma(\gamma^+ \phi \to {\rm else})$ contains at least one heavy particle (in the UV theory) in the final state and corresponds to elastic amplitudes generated by higher dimensional operators.  Here, we only have one contribution to $\sigma(\gamma^+ \phi \to {\rm 2~light particles})$, which is $\sigma(\gamma^+ \phi \to \gamma \phi)$ where the final state photon polarization is summed over.  The key observation is that, in the UV theory, the tree-level cross section $\sigma(\gamma^+ \phi \to \gamma \phi)$ has no contribution from heavy particles due to the structure of the gauge interaction, which could not mix light and heavy particles.  This is illustrated in \autoref{fig:apsigma}.  
As an result, $\sigma(\gamma^+ \phi \to \gamma \phi)$ contains only dim-4 contributions and the corresponding part can be subtracted from the dispersion relation, while the remaining part (which we denote as $\Sigma'$) must be positive since $\sigma(\gamma^+ \phi \to {\rm else})$ is positive:   
\begin{equation}
    \begin{split}
        \Sigma^{\prime} \equiv \Sigma - \frac{ 2 }{ \pi } \int_{0}^{\infty} ds \frac{ s^{4} \sigma( \gamma^{+} \phi \to \gamma \phi ) }{ ( s^{2} + s_{0}^{2} )^{3} }
        = \frac{ 2 }{ \pi } \int_{0}^{\infty} ds \frac{ s^{4} \sigma( \gamma^{+} \phi \to {\rm else} ) }{ ( s^{2} + s_{0}^{2} )^{3} }
        \geq 0 \, .
    \end{split} \label{eq:sigmapxs}
\end{equation}
More explicitly, we have
\begin{equation}
    \sigma( \gamma^{+} \phi \to \gamma \phi ) = \frac{ e^{4} }{ 4 \pi s } \, \hspace{0.5cm}  \Longrightarrow \hspace{0.6cm} \frac{2}{\pi} \int_0^{\infty} ds\,\frac{ s^4 \sigma( \gamma^{+} \phi \to \gamma \phi ) }{\left(s^2+s_0^2\right)^3} = \frac{ e^{4} }{ 8 \pi^2 } \frac{ 1 }{ s_{0}^{2} } \, ,
\end{equation}
which is exactly the dim-4 contribution in \autoref{eq:phi-gamma-sigma}.  From now on we will subtract this term and work with $\Sigma'$ instead.

While \autoref{eq:phi-gamma-sigma} is absent of any UV divergences (which are treated with the usual renormalization procedure, {\it i.e.}, canceled by counter terms), it contains IR poles in the form $1/\epsilon$ and $1/\epsilon^2$, which comes from the master integrals $C_0$ and $D_0$.  This is expected since the cancellation of these IR poles require the real-emission diagrams which we do not include.  The treatment of these IR poles in the dispersion relation is a nontrivial issue in general~\cite{Bellazzini:2020cot,Chang:2025cxc,Beadle:2025cdx,Desai:2025alt}.  However, it turns out that, once we use the assumption that the coefficients in \autoref{eq:c1loop}, except $c_{D^{4}\phi^{4}}^{(1)}$ and $c_{D^{4}\phi^{4}}^{(2)}$, are generated at the one-loop level, we can discard the contributions in \autoref{eq:phi-gamma-sigma} that are at least generated at the two-loop order in the UV theory. This removes all the IR poles, and $\Sigma'$ now becomes  
\begin{equation}
    \label{eq: sigma log t}
    \begin{split}
    \Sigma^{\prime} \underset{t\to 0}{=}
    - \frac{1}{\Lambda^{4}} 
    \frac{1}{2} c_{F^{2}D^{2}\phi^{2}}^{(1)}+\frac{1}{\Lambda^4}\frac{19 e^2}{576\pi^2}(2 c_{D^4\phi^4}^{(1)} + c_{D^4\phi^4}^{(2)})-\frac{1}{\Lambda^4}\frac{e^2}{96\pi^2}(2 c_{D^4\phi^4}^{(1)} +  c_{D^4\phi^4}^{(2)})\log \frac{-t}{\mu^2}
    \,,
    \end{split}
\end{equation}
which contains the tree-level contribution of the one-loop generated coefficient $c_{F^{2}D^{2}\phi^{2}}^{(1)}$, and the one-loop contribution of $c_{D^4\phi^4}^{(1)}$ and $c_{D^4\phi^4}^{(2)}$, which we now assume to be tree-level generated.\footnote{More precisely, we assume the combination $2 c_{D^4\phi^4}^{(1)} + c_{D^4\phi^4}^{(2)}$ is generated at the tree level.  For instance, we can have $c_{D^4\phi^4}^{(1)}$ tree-level generated and $c_{D^4\phi^4}^{(2)}$ loop generated, as in Model~II in \autoref{subsec:topdow}.}  

\begin{figure}[t]
	\centering
	\includegraphics[width=0.75\textwidth]{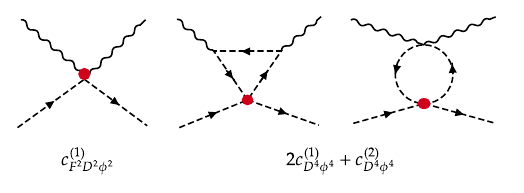}
	\caption{The corresponding Feynman diagrams for \autoref{eq: sigma log t} and \autoref{eq: sigma log m}.  The tree-level contribution comes from $c_{F^{2}D^{2}\phi^{2}}^{(1)}$. The one-loop diagrams are proportional to the combination  $2 c_{D^4\phi^4}^{(1)} + c_{D^4\phi^4}^{(2)}$.  Additional one-loop diagrams which contain vanishing scaleless integrals (in the massless case) are not included here.   
    }
    \label{fig:ap-1loop}
\end{figure}

It is peculiar that the $\Sigma'$ in \autoref{eq: sigma log t} has no $s_0$ dependence.  This can be understood from the loop structure, as shown in \autoref{fig:ap-1loop}.   
In particular, only the Mandelstam variable $t$ enters the loops generated by $c_{D^4\phi^4}^{(1)}$ and $c_{D^4\phi^4}^{(2)}$ (see the master integrals in \autoref{sec:masint}), resulting in the $\log (-t)$ dependence in \autoref{eq: sigma log t}.  As such, $\Sigma'$ still contain IR divergence in the forward limit ($t\to 0$), which needs to regulated.  
To do so, we introduce a small mass $m$ for the scalar field $\phi$, and calculate the one-loop amplitude from $c_{D^4\phi^4}^{(1)}$ and $c_{D^4\phi^4}^{(2)}$, which now becomes
\begin{equation}
\label{eq: amp with mass}
\mathcal{A}^{[8],\text{1-loop}}_{+ \phi \to + \phi} \approx \frac{e^2}{96\pi^2} s (s+t) (2 c_{D^4\phi^4}^{(1)} + c_{D^4 \phi^4}^{(2)})\left(2 - \log \left(\frac{m^2-t}{\mu^2}\right)\right) + \mathcal{O}(m^2)\,,
\end{equation}
where we have done an expansion in $m^2$ and kept only the leading $\log m^2$ dependence.
One could then safely take the forward limit, and \autoref{eq: sigma log t} becomes  
\begin{equation}
    \label{eq: sigma log m}
    \begin{split}
    \Sigma^{\prime} \cdot 2 \Lambda^4 \,  \approx& 
    -  c_{F^{2}D^{2}\phi^{2}}^{(1)}+\frac{19 e^2}{288\pi^2}( 2c_{D^4\phi^4}^{(1)} + c_{D^4\phi^4}^{(2)})-\frac{e^2}{48\pi^2}( 2c_{D^4\phi^4}^{(1)} + c_{D^4\phi^4}^{(2)})\log \frac{m^2}{\mu^2}
    \geq 0 \,,
    \end{split}
\end{equation}
where we have also multiplied both sides by $2\Lambda^4$ for convenience.  
Essentially, we have replaced the $-t$ by $m^2$ in the log term.  \autoref{eq: sigma log m} is now well defined and we can study its implications.  However, it should be emphasized that $m^2$ is a regulator and its value should always be taken to be much smaller than any finite scale in the theory, such as $s_0$ or $\Lambda$. \footnote{We also note that in the recent studies Refs.~\cite{Caron-Huot:2021rmr,Caron-Huot:2022ugt,Beadle:2024hqg,Beadle:2025cdx,Chang:2025cxc}, a different approach has been taken to avoid the forward-limit divergence, which involves considering the dispersion relation at finite $t$ and integrating it over with some smearing function $f(t)$. It would be interesting to see if this approach would provide additional useful information.  This is however beyond the scope of our current study.}

The absence of additional IR divergence in \autoref{eq: sigma log m} can also be understood with the following argument. 
Since the optical theorem is valid order-by-order in the perturbation theory, so is the dispersion relation in \autoref{eq:Sigmapos}.
As mentioned above \autoref{eq:cs2to2}, for the one-loop forward elastic amplitudes we consider, the corresponding total cross section (via the optical theorem) is a tree-level 2-to-2 one.  Such tree-level cross sections should not have any $1/\epsilon$ dependence associated with dimensional regularization (with $D = 4 - 2 \epsilon$).  Therefore, the LHS of the dispersion relation in \autoref{eq:Sigmapos} should also be free from such divergences.  
On the other hand, the $1/\epsilon$ dependent terms in \autoref{eq:phi-gamma-sigma} are formally at the two-loop level in the UV theory, and the corresponding cross section does contain loop contributions (depending on how one cuts the diagram), which contain IR divergences that are only canceled when summed with real emission diagrams, according to the renowned Kinoshita–Lee–Nauenberg (KLN) theorem~\cite{Kinoshita:1962ur, Lee:1964is}.  
Note that these arguments should also be applicable to a broad range of EFTs.  
To obtain positivity bounds beyond the one-loop level, one may need to develop a systematic method to add the real emission diagrams and cancel the IR divergences, which is beyond the scope of our current study.  On the other hand, at the one-loop level, we do not expect the positivity bounds to be spoiled by any divergences at the two-loop level, which must be canceled by other contributions at the same (two-loop) order.

It is not a coincidence that only the combination $2 c_{D^4\phi^4}^{(1)} + c_{D^4\phi^4}^{(2)}$ enters \autoref{eq: sigma log m}, since it is also the same combination that enters the forward elastic $\phi\phi \to \phi\phi$ amplitude, with the corresponding tree-level positivity bound being\footnote{Note that $\mathcal{A}_{\phi \phi \to \phi \phi}$ is not $s\leftrightarrow u$ symmetric, but the tree-level bound can be obtained with other (simpler) versions of dispersion relations.}  
\begin{equation}
\label{eq:phi-phi-treebound}
    2 c_{D^4\phi^4}^{(1)} + c_{D^4\phi^4}^{(2)} \geq 0 \,.
\end{equation}
Again, here we assume the combination $2 c_{D^4\phi^4}^{(1)} + c_{D^4\phi^4}^{(2)}$ is generated at the tree-level so this bound holds for \autoref{eq: sigma log m}.  Then, assuming $\mu^2$ is not too small ({\it i.e.} $\mu^2 \gg m^2$), the last term in \autoref{eq: sigma log m} is positive and could easily dominate over the first two terms.  As such, \autoref{eq: sigma log m} no longer implies the tree-level bound $c_{F^{2}D^{2}\phi^{2}}^{(1)} \leq 0$.  In \autoref{subsec:topdow} we will show examples where indeed the one-loop generated $c_{F^{2}D^{2}\phi^{2}}^{(1)}$ could violate the tree-level bound. 

It is illustrative to look at the $\beta$-function of $c_{F^{2}D^{2}\phi^{2}}^{(1)}$.  The general expression in \autoref{eq:beta1} now becomes 
\beq
\label{eq:beta2}
\beta(c^{(1)}_{F^2 D^2\phi^2}) = \frac{e^2}{24 \pi^2} \left( 2 c_{D^4 \phi^4}^{(1)} + c_{D^4\phi^4}^{(2)}\right)\,, 
\eeq 
which can also be obtained by simply taking \autoref{eq: sigma log m} and requiring $\mu \frac{d}{d\mu} \Sigma' =0$.  Indeed, both the amplitude and $\Sigma'$ must be independent of the unphysical renormalization scale $\mu$, which is exactly the condition that gives the $\beta$-function of $c_{F^{2}D^{2}\phi^{2}}^{(1)}$.  Here, by combining \autoref{eq:phi-phi-treebound} and \autoref{eq:beta2} we obviously have $\beta(c^{(1)}_{F^2 D^2\phi^2}) \geq 0$.  However, suppose the exact form of $\beta(c^{(1)}_{F^2 D^2\phi^2})$ is unknown, one could write \autoref{eq: sigma log m} in the more general form
\begin{equation}
    \label{eq: sigma log m beta}
    \begin{split}
    \Sigma^{\prime} \cdot 2 \Lambda^4 \,  \approx& 
    -  c_{F^{2}D^{2}\phi^{2}}^{(1)}-\frac{\beta}{2}\log \frac{m^2}{\mu^2} + \mbox{ finite 1-loop terms}  \,,
    \end{split}
\end{equation}
where, assuming the only physical scale in the log term is $m^2$, this is the only possible form as required by $\mu \frac{d}{d\mu} \Sigma' =0$.  Given that $m^2$ is a regulator that can be taken as infinitesimal, and assuming  $ c_{F^{2}D^{2}\phi^{2}}^{(1)}$ is at least finite for some finite $\mu^2$, we can conclude that $\Sigma'\geq 0$ implies $\beta(c^{(1)}_{F^2 D^2\phi^2}) \geq 0$.  
This means that, while $c_{F^{2}D^{2}\phi^{2}}^{(1)}$ does not necessarily satisfy the tree-level bound  $c_{F^{2}D^{2}\phi^{2}}^{(1)}\leq 0$, its $\beta$-function always tend to restore the tree-level bound in the IR, so $c_{F^{2}D^{2}\phi^{2}}^{(1)}\leq 0$ is always satisfied for sufficiently small $\mu$ (assuming the perturbation theory is still valid).    
However, it should be emphasized that by changing the value $\mu$ we do not obtain new positivity bounds --- it merely changes the definition of couplings, and for small $\mu$ we are simply absorbing a large positively-bounded log term in the definition of $ c_{F^{2}D^{2}\phi^{2}}^{(1)}$.

The positivity of the $\beta$-function is also related to the fact that the one-loop diagrams in \autoref{fig:ap-1loop} are in a ``symmetric'' form, where the corresponding diagrams in the UV corresponds to total square terms in the cross section via the optical theorem, which are naturally bounded to one side~\cite{Ye:2024rzr}. 
Indeed, it would be interesting to understand more generally what conditions are needed for $\beta$-functions to be bounded to one-side, and formulate some kind of ``EFT $a$-theorem'' similar to the one for the dim-6-squared contribution in forward elastic amplitudes proposed in Ref.~\cite{Liao:2025npz}.

Finally, we also note that when $c_{D^4\phi^4}^{(1)}$ and $c_{D^4\phi^4}^{(2)}$ are also loop-generated, their contribution to \autoref{eq: sigma log m} becomes sub-leading and can be ignored. As a result, the $\Sigma^{\prime}$ bound reduces to the tree-level positivity bound $c_{F^{2}D^{2}\phi^{2}}^{(1)} < 0$, which remains valid.  In this case, we could not determine the sign of $\beta(c^{(1)}_{F^2 D^2\phi^2})$ (given by the more general expression \autoref{eq:beta1}) from the dispersion relation.


\subsection{Top-down perspectives from UV models}
\label{subsec:topdow}

In this section, we will take a top-down perspective and try to verify our results in specific UV models.  
Functional methods are employed to perform matching beyond the tree level, as outlined in previous works~\cite{Henning:2014wua, Henning:2016lyp, Cohen:2020fcu}. Specifically, for a given UV model, after obtaining the one-light-particle-irreducible (1LPI) effective action $\Gamma_{\text{L}}[\phi]$ by integrating out the heavy fields, the Wilson coefficients can be identified through the following matching condition:
\begin{equation}
    \Gamma_{ \text{L,EFT} }\left(c_{i}^{[j]}, \mu = \mu_m\right) = \Gamma_{ \text{L,UV} }\left(g,\mu=\mu_m\right) \,,
\end{equation}
where $ \mu_m $ is the matching scale. It is important to note that the UV theory is assumed to be weakly coupled around the matching scale in this framework. In the present work, one-loop matching is carried out using the \texttt{Matchete} package~\cite{Fuentes-Martin:2022jrf}.  

After matching, the Wilson coefficients could then be run down to a lower scale with the renormalization group equations (RGE) in the EFT.  
The leading order solutions of the RGE are of the form
\begin{equation}
    c_{i}^{[j]}(\mu) = c_{i}^{[j]}(\mu_m) + \beta( c_{i}^{[j]}(\mu_m) ) \log \frac{\mu}{\mu_m} \,, \label{eq:crun1}
\end{equation}
where the index $[j]$ denotes the corresponding operator dimension.  The general form of the dim-8 $\beta$-function is given by
\begin{equation}
    \beta(c_i^{[8]}) = \frac{1}{16 \pi^2} \left( \sum_{j} \gamma_{ij} c_{j}^{[8]} + \sum_{jk} \gamma^{\prime}_{ijk} c_{j}^{[6]} c_{k}^{[6]}  \right) \,,
\end{equation}
where $\gamma_{ij}$ and $\gamma^{\prime}_{ijk}$ are the anomalous dimension matrices.
Note that when $\log(\mu / \mu_m)$ is very large, \autoref{eq:crun1} becomes invalid and a resummation of the large logs is necessary.

We will consider two different UV model.  The first model is somewhat trivial where all higher dimensional operators are generated at the one-loop level.  In this case, the leading contribution to $\Sigma'$ still comes from the tree-level (in the EFT) contribution of $c_{F^{2}D^{2}\phi^{2}}^{(1)}$, and its tree-level bound is still valid.  
In the second model, only the operator coefficient $c_{D^{4}\phi^{4}}^{(1)}$ is generated at the tree level.  In this case, the dispersion relation in \autoref{eq: sigma log m} has a non-trivial implication.  
For convenience, we will also identify the cutoff scale $\Lambda$ with the mass of the heavy particle $M$ for the rest of this section, and simply set $\Lambda=M$.

\subsubsection*{Model~I.}
In this model we introduce a heavy complex field $\Phi$ with charge $ +1/2$.  The Lagrangian is given by
\begin{equation}
\label{eq:testUV1}
    \begin{split}
        \mathcal{L}_{\uppercase\expandafter{\romannumeral1}} =& - \frac{1}{4} F^{\mu \nu} F_{\mu \nu} + ( D^{\mu} \phi )^{\dagger} ( D_{\mu} \phi ) + ( D^{\mu} \Phi )^{\dagger} ( D_{\mu} \Phi ) -M^2\Phi^{\dagger}\Phi \\ 
        &- \frac{1}{4} \lambda_{1} ( \phi^{\dagger} \phi )^{2} - \frac{1}{4} \lambda_{2} ( \Phi^{\dagger} \Phi )^{2} - \lambda_3\Phi^{\dagger}\Phi\phi^{\dagger}\phi -( \frac{1}{2} g M \Phi^{\dagger} \Phi^{\dagger} \phi + h.c. )\,.
    \end{split}
\end{equation}
Since all interactions of $\Phi$ contain at least two $\Phi$ (or $\Phi^\dagger$) legs, no amplitudes in the EFT can be UV completed with a tree-level exchange of $\Phi$.  All operators in \autoref{eq:La6} and \autoref{eq:La8} are generated at the one-loop level.  In particular,  we have  
\begin{equation}
\label{eq:match1-1}
    c_{F^{2}D^{2}\phi^{2}}^{(1)} = - \frac{1}{16\pi^2} \frac{1}{2520} g^2 e^2 \,,
\end{equation}
and its $\beta$-function (as in \autoref{eq:beta1}) also vanishes at the one-loop level since all the coefficients on the RHS of \autoref{eq:beta1} are already generated at the one-loop level, making it a two-loop effect in the UV model.\footnote{This could also be understood from the fact that in this model, the one-loop diagrams to the $\apap$ amplitude involve only the heavy particle $\Phi$ in the loop.}  
The complete one-loop matching and (no) running results can be found in \autoref{eq:match1-2}.  In this case, the leading contribution to $\Sigma'$ is
\begin{equation}
\label{eq: sigma model1}
\Sigma' \approx -\frac{1}{M^4} \frac{1}{2} c_{F^2 D^2 \phi^2}^{(1)}   \geq  0 \, ,
\end{equation}
which is obviously satisfied by \autoref{eq:match1-1}.  It is interesting to note that $c_{F^{2}D^{2}\phi^{2}}^{(1)}$ has no contribution from the quartic coupling $\lambda_3$ --- there is a  diagram with one insertion of $\lambda_3$ for the one-loop $\apap$ amplitude, but it only contribute to $c_{F^{2}D^{2}\phi^{2}}^{(2)}$ (see \autoref{eq:match1-2}).  This is consistent with the positivity bound since a $\lambda_3$ contribution could have taken either sign and easily violated the bound.  
With the UV model we could also verify the dispersion relation by explicitly calculating the cross section contribution on the RHS of \autoref{eq:Sigmapos}. After subtracting the pure dim-4 contribution as in \autoref{eq:sigmapxs}, the only cross section left is $\sigma(\gamma^{+} \phi \to \Phi \Phi)$, which is
\begin{equation}
\label{eq:csUV1}
    \begin{split}
        \sigma( \gamma^{+} \phi \to \Phi \Phi ) =&~ \frac{ 1 }{ 32 \pi s^3 } e^2 g^2 M^2 \Bigg\{ (2M^2 + s) \log \left( \frac{ \sqrt{s} + \sqrt{ s - 4M^2 } }{ \sqrt{s} - \sqrt{ s - 4M^2 } } \right)\\
        &- 3 \sqrt{ s (s-4M^2) } \Bigg\} \Theta( s - 4M^2 ) \, ,
    \end{split}
\end{equation}
where $\Theta(s - 4 M^{2})$ is a Heaviside step function. 
We then have 
\begin{equation}
\label{eq:csSiUV1}
    \Sigma^{\prime} = \frac{2}{\pi} \int_0^{\infty} ds\,\frac{ s^4 \sigma( \gamma^{+} \phi \to \Phi \Phi ) }{\left(s^2+s_0^2\right)^3} = \frac{ g^2 e^2 }{ M^4 } \frac{ 1 }{ 80640 \pi^2 } + \mathcal{O} \left( \frac{ s_0 }{ M^6 } \right) \, ,
\end{equation}
which agrees exactly with \autoref{eq: sigma model1} with \autoref{eq:match1-1} substituted in.

We also note that, since all higher dimensional operators are generated at the one-loop level in this model, the full dispersion relation in \autoref{eq:phi-gamma-sigma} are valid up to the two-loop level in the UV model.  However, its implication becomes unclear beyond the one-loop level.  First, at the two-loop level in the UV model, heavy particle loops can clearly contribute to $\sigma(\gamma^+\phi \to \gamma \phi)$, so in general one could not subtract the dim-4 term $\frac{e^4}{8 \pi^2} \frac{ 1 }{ s_{0}^{2} }$ from \autoref{eq:phi-gamma-sigma} and still claim the rest is positive.  Second, \autoref{eq:phi-gamma-sigma} would also contain the problematic IR divergences as discussed in \autoref{subsec:Imp}.

\subsubsection*{Model~II.}
In this model we introduce a heavy complex scalar $\Phi$ with charge 2, which makes it possible to write down a $\Phi^\dagger \phi\phi$ (or $\Phi \phi^\dagger \phi^\dagger$) trilinear coupling.  The full Lagrangian is given by
\begin{equation}
\label{eq:testUV2}
    \begin{split}
        \mathcal{L}_{\uppercase\expandafter{\romannumeral2}} =& - \frac{1}{4} F^{\mu \nu} F_{\mu \nu} + ( D^{\mu} \phi )^{\dagger} ( D_{\mu} \phi ) + ( D^{\mu} \Phi )^{\dagger} ( D_{\mu} \Phi ) -M^2\Phi^{\dagger}\Phi \\ 
        &- \frac{1}{4} \lambda_{1} ( \phi^{\dagger} \phi )^{2} - \frac{1}{4} \lambda_{2} ( \Phi^{\dagger} \Phi )^{2} - \lambda_3\Phi^{\dagger}\Phi\phi^{\dagger}\phi -( \frac{1}{2} g M \Phi^{\dagger} \phi \phi + h.c. )\,.
    \end{split}
\end{equation}
Here, the scalar trilinear coupling could generate $c_{D^{4}\phi^{4}}^{(1)}$ at the tree level, while all other coefficients that contribute to the dispersion relation in \autoref{eq:phi-gamma-sigma} are generated at the one-loop level.  The complete one-loop matching results are given by \autoref{eq:match2-2}. 
In particular, at the matching scale we have 
\begin{equation}
\label{eq:match2-1-M}
    c_{F^{2}D^{2}\phi^{2}}^{(1)} (\mu=M) = \frac{1}{16\pi^2} \frac{3}{2} e^2 g^2  \,.
\end{equation}
Note that, just as in Model~I, here $c_{F^{2}D^{2}\phi^{2}}^{(1)}$ is still independent of the quartic coupling $\lambda_3$. For its one-loop $\beta$-function in \autoref{eq:beta2}, we only need to include the tree-level matching results for $c_{D^4 \phi^4}^{(1)}$ and $c_{D^4 \phi^4}^{(2)}$, which are
\beq
c_{D^4 \phi^4}^{(1)} = g^2 \,, \quad\quad c_{D^4 \phi^4}^{(2)} = 0 \,.  \label{eq:md2cdphi}
\eeq
We then have
\beq
\beta(c^{(1)}_{F^2 D^2\phi^2}) = \frac{1}{12\pi^2}e^2 g^2 \,,
\eeq
which gives, at the leading order (without resummation of logs)
\begin{equation}
\label{eq:match2-1}
    c_{F^{2}D^{2}\phi^{2}}^{(1)} = \frac{1}{16\pi^2} \frac{1}{6} e^2 g^2  \left(9 + 4 \log \left(\frac{\mu^2}{M^2}\right)\right)\,.
\end{equation}
Indeed, we now see that the tree-level bound $c_{F^{2}D^{2}\phi^{2}}^{(1)} \leq 0$ is violated by \autoref{eq:match2-1} for $\mu \gtrsim 0.3 M$, while the positive $\beta$-function ensures that the tree-level bound is restored for sufficiently small $\mu$, consistent with the results in \autoref{subsec:Imp}.   

For the dispersion relation in \autoref{eq: sigma log m} we also need the one-loop result of $ c_{F^{2}D^{2}\phi^{2}}^{(1)}$ and the tree-level results for $c_{D^4 \phi^4}^{(1)}$ and $c_{D^4 \phi^4}^{(2)}$.  After plugging in \autoref{eq:match2-1} and \autoref{eq:md2cdphi}, it becomes
\begin{equation}
\label{eq: log Sigma with mass}
\Sigma^{\prime} = \frac{e^2 g^2}{576 M^4 \pi^2}\left( 11-12\log\left(\frac{m^2}{M^2}\right) \right) \,,
\end{equation}
which is positive (since we always have $m \ll M$) and independent of $\mu^2$ as expected.  
One could also compute $\Sigma'$ from the cross section in this model.  In this case, the only contribution is from $\sigma(\gamma^+ \phi \to \Phi \phi^{\dagger})$.  When $\phi$ is massless, the differential cross section with respect to the production polar angle $\theta$ (between the outgoing $\Phi$ and incoming $\gamma$) is given by 
\begin{equation}
\label{eq: diff cross section}
    \frac{ d \sigma(\gamma^+ \phi \to \Phi \phi^{\dagger}) }{ d \cos \theta } = \frac{ e^2 g^2 M^2}{16 \pi} \frac{ (1 - \cos \theta) (s - M^2) \left( M^2(1+\cos\theta) - s(3+\cos\theta) \right)^2 }{ s^3 (1 + \cos \theta) \left( M^2(1+\cos\theta) + s(1-\cos\theta) \right)^2 } \,,
\end{equation}
which diverges as $\theta \to \pi$. This divergence can be regulated by adjusting the integration domain of $ d\cos \theta$ from $ [-1, 1] $ to $ [-1 + 2 m^2 / s, 1] $, which is equivalent to introducing a small mass $ m $ for the scalar field $ \phi $.\footnote{When introducing a tiny mass $m$, the singular term $1 + \cos \theta$ in the denominator of differential cross section becomes $ 1 + \cos \theta \sqrt{ 1 - 4 m^2 s / ( s - M^2 + m^2 )^2 } \approx 1 + \cos \theta ( 1 - 2 m^2 / s )$, and we change the variable to $\cos \tilde{\theta} = \cos{\theta} ( 1 - 2 m^2 / s )$, so the integration domain become $[-1 + 2 m^2 / s, 1 - 2 m^2 / s]$.  The correction to the upper range has no impact in the $m\to 0$ limit since there is no divergence at $\cos\theta\to 1$. } After that, the total cross section is finite and given by:
\begin{equation}
\sigma(\gamma^+ \phi \to \Phi \phi^{\dagger}) \approx 
\frac{e^2 g^2 M^2}{8 \pi s^3} \left(4 s \log\left(\frac{s}{M^2}\right)-(s-M^2) \left(\log\left(\frac{m^2}{s}\right)+5\right) \right) \Theta(s-M^2)\,.
\end{equation} 
When taking $m \to 0$ limit, the cross section becomes logarithmic divergent. From the RHS of the dispersion relation we then have 
\begin{equation}
\label{eq: log cross section}
\Sigma' = \frac{2}{\pi} \int_0^{\infty} ds\,\frac{ s^4 \sigma(\gamma^+ \phi \to \Phi \phi^{\dagger}) }{\left(s^2+s_0^2\right)^3} = \frac{e^2 g^2}{576 M^4 \pi^2} \left(11-12\log\left(\frac{m^2}{M^2}\right)\right) + 
\mathcal{O} \left( \frac{ s_0 }{ M^6 } \right)\,,
\end{equation}
in agreement with \autoref{eq: log Sigma with mass}.

It should be noted that, when $\mu \ll M$, the leading order result in \autoref{eq:match2-1} is invalid, and one needs to resum the large logs.  This in general involves solving a set of coupled differential equations, which can be difficult to do analytically.  On the other hand, the $\Sigma'$ in \autoref{eq: log Sigma with mass} explicitly contains a very large $\log (m^2/M^2)$, where $m$ is a small regulator and $M$ is the cutoff scale of the EFT.      
As already pointed out in Ref.~\cite{Ye:2024rzr}, it is likely very challenging (if possible at all) to perform the resummation at the amplitude level and obtain a resumed dispersion relation, since the kinematic dependence is very complicated for higher loop orders.  
Nevertheless, since $\Sigma'$ is not an observable, it is not particularly problematic for it to contain a large log term.  
As mentioned above, we expect the dispersion relation \autoref{eq:Sigmapos} to hold order-by-order,
despite the large log term --- in fact, we have just verified that it holds at the leading order here with \autoref{eq: log Sigma with mass} and \autoref{eq: log cross section}.


\section{Photon-photon scattering}
\label{sec:phopho}

We now consider the $\aaaa$ amplitude.  Here, the two incoming photons can take different polarizations, while elasticity requires the final state polarizations to be the same as the initial ones.  
While all polarizations can be considered, we will focus on linear polarizations which are invariant under a $s \leftrightarrow u$ crossing, as required by our dispersion relation \autoref{eq:Sigmapos}.\footnote{On the other hand, helicity states are flipped under the  $s \leftrightarrow u$ crossing. See also discussions in Refs.\cite{Bellazzini:2015cra,Bellazzini:2016xrt,Remmen:2019cyz}.} 
We choose the basis polarization vectors as $\epsilon^{x}$ and $\epsilon^{y}$, which are the real linear polarizations along the $x$ and $y$ axes, respectively.  They are related to the helicity polarization vectors by
\begin{equation}
    \epsilon^{x}(p_{i}) = \frac{ 1 }{ \sqrt{2} } \Big( \epsilon^{+}(p_{i}) + \epsilon^{-}(p_{i}) \Big), \quad \epsilon^{y}(p_{i}) =\frac{ 1 }{ \sqrt{2} i } \Big( \epsilon^{+}(p_{i}) - \epsilon^{-}(p_{i}) \Big) \,.
\end{equation}

We start with the tree-level EFT amplitudes, for which only the coefficients $c_{F^{4}}^{(1)}$ and $c_{F^{4}}^{(2)}$ could contribute.  There is also no dim-4 or dim-6 contribution at the tree level.   
For instance, we can consider the amplitude $\gamma^{y}\gamma^{y} \to \gamma^{y}\gamma^{y}$, where the two incoming photons have the same polarization.  Its tree-level amplitude is symmetric under $s$, $t$ and $u$, given by 
\begin{equation}
\label{eq:4-gamma-treeamp-1}
    \mathcal{A}^{[8],\text{tree}}_{yy \to yy} = \frac{ 8 }{ \Lambda^{4} } c_{F^{4}}^{(1)} (s^2 + t^2 + u^2) \,.
\end{equation}
Taking the forward limit ($t\to 0$) and applying \autoref{eq:Sigmapos} yields the tree-level bound
\begin{equation}
\label{eq:4gamma-treebound-1}
    c_{F^{4}}^{(1)} \geq 0 \,.
\end{equation}
We could also consider the case that the two incoming photons have orthogonal polarizations, $\gamma^{y} \gamma^{x} \to \gamma^{y} \gamma^{x}$. 
The tree-level amplitude is given by
\begin{equation}
\label{eq:4-gamma-treeamp-2}
    \mathcal{A}^{[8],\text{tree}}_{y x \to y x} = - \frac{ 8 }{ \Lambda^{4} } \left( 2 c_{F^{4}}^{(2)} \, su + ( c_{F^{4}}^{(1)} - c_{F^{4}}^{(2)} ) \, t^2 \right) \,,
\end{equation}
which clearly also satisfies the crossing symmetry $s \leftrightarrow u$.  Taking the $t\to 0$ limit, it gives the bound
\begin{equation}
\label{eq:4gamma-treebound-2}
    c_{F^{4}}^{(2)} \geq 0 \,.  
\end{equation}
The same bounds could also be derived by considering the propagation of photon in background fields and requiring the photon to be non-superluminal, as pointed out in Ref.~\cite{Adams:2006sv}.

\begin{figure}[t]
    \centering
    \includegraphics[width=0.6\linewidth]{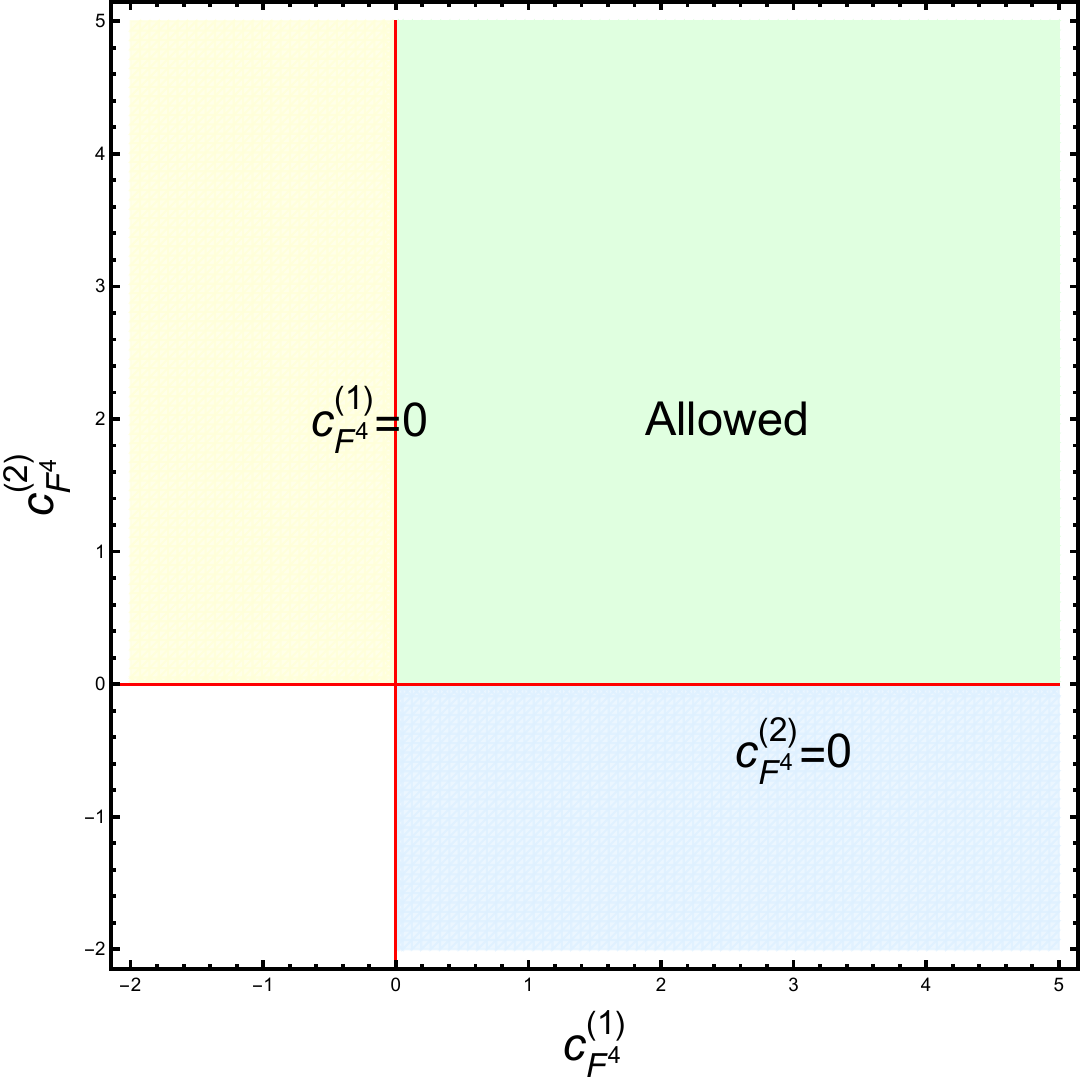}
    \caption{The positivity bound of $c_{F^4}^{(1)}$ and $c_{F^4}^{(2)}$. The red lines stand for $c_{F^4}^{(1)}=0$ and $c_{F^4}^{(2)}=0$. The green region bounded by $c_{F^4}^{(1)}\geq 0$ and $c_{F^4}^{(2)}\geq 0$ already satisfies the more general set of bounds given by \autoref{eq:cf4posgen}.     
    }
    \label{fig: four gamma bound}
\end{figure}

While other combinations of polarizations can also be considered, it turns out that the combination of \autoref{eq:4gamma-treebound-2} and \autoref{eq:4gamma-treebound-1} already gives the best bound.  More specifically, we can obtain a continuous set of positivity bounds associated with the scattering of two photons with arbitrary linear polarizations $\epsilon^{\alpha_{1,2}}(p_{1,2}) = \epsilon^{x} (p_{1,2}) \cos \alpha_{1,2} + \epsilon^{y} (p_{1,2}) \sin \alpha_{1,2}$, which are
\begin{equation}
\label{eq:cf4posgen}
    \begin{split}
        ( c_{F^{4}}^{(1)} - c_{F^{4}}^{(2)} ) \cos( 2 (\alpha_{1} - \alpha_{2}) ) + c_{F^{4}}^{(1)} + c_{F^{4}}^{(2)} \geq 0 \,.
    \end{split}
\end{equation}
The intersection of this continuous set of bounds is exactly the region carved out by the two bounds $c_{F^2}^{(1)}\geq 0$ and $c_{F^4}^{2}\geq 0$, as shown in \autoref{fig: four gamma bound}. 
These bounds (as well as additional ones involving $d>8$ operator coefficients) have already been pointed out in many previous studies such as Refs.~\cite{Bellazzini:2019xts,Remmen:2019cyz,Henriksson:2021ymi}.

We will now compute the $\aaaa$ amplitudes at the one-loop level in the EFT, and derive the corresponding dispersion relations, which we will denote as $\Sigma_{yy}$ and $\Sigma_{yx}$ for the two polarization configurations.    
First of all, we note that, the one-loop contributions to the $\aaaa$ amplitudes in any UV model involves either the light scalar $\phi$, or a heavy particle, {\it i.e.} the one loop diagram could not mix heavy and light particles.  As a result, at the one-loop level in the UV model, the $\aaaa$ amplitudes only receive the contribution from the one-loop generated $c_{F^{4}}^{(1)}$ and $c_{F^{4}}^{(2)}$, in which case the tree-level bounds \autoref{eq:4gamma-treebound-2} and  \autoref{eq:4gamma-treebound-1} still hold.  Nevertheless, the dispersion relations still contain useful information, as we will show below.

Starting with $\gamma^{y} \gamma^{y} \to \gamma^{y} \gamma^{y}$, the calculation of the dim-4 one-loop amplitude and its contribution to the dispersion relation in \autoref{eq:Sigmapos} contain some subtleties that are discussed in details in \autoref{sec:aaaadim4}.  However, just like the case for $\apap$, here the one-loop dim-4 contribution corresponds to the 2-to-2 tree-level cross section $\sigma(\gamma^{y} \gamma^{y} \to \phi \phi^\dagger)$ on the RHS of \autoref{eq:Sigmapos}, which receives no contribution from heavy particles and can be subtracted from $\Sigma_{yy}$, with
\begin{equation}
    \begin{split}
        \Sigma^{\prime}_{yy} \equiv \Sigma_{yy} - \frac{ 2 }{ \pi } \int_{0}^{\infty} ds \frac{ s^{4} \sigma(\gamma^{y} \gamma^{y} \to \phi \phi^\dagger) }{ ( s^{2} + s_{0}^{2} )^{3} }        
        \geq 0 \, ,
    \end{split} \label{eq:sigmaa1xs}
\end{equation}
where $\Sigma^{\prime}_{yy}$ still contains all the contributions from higher dimensional operators.  
It also turns out that the dim-6 one-loop contribution vanishes as a result of the $stu$ full crossing symmetry,\footnote{Note again that the dim-6 one-loop amplitude contains no log terms since there is no tree-level dim-6 contribution, so the dim-6 one-loop contribution is proportional to a linear combination of $s$, $t$ and $u$, which vanishes due to the full crossing symmetry and the relation $s+t+u=0$.} so $\Sigma^{\prime}_{yy}$ contain only dim-8 (or dim-6 squared) contributions.  
The dim-8 one-loop amplitude is given by
\beq
\label{eq:4-gamma-loopamp-1}
\begin{split}
\mathcal{A}^{[8],\text{1-loop}}_{yy \to yy} = & \frac{1}{\Lambda^4} \frac{1}{96 \pi^2} \Big[ 24 c_{F^2 \phi^2}^{2} ( s^2 B_0(s) + t^2 B_0(t) + u^2 B_0(u) )  \\
    &- e^2 c_{F^2 D^2 \phi^2}^{(1)} \Big( (t^2 + u^2) B_0(s) + (s^2 + u^2) B_0(t) + (s^2 + t^2) B_0(u) \Big) \Big]  \\
    &-\frac{1}{\Lambda^4} \frac{ 1 }{288 \pi^2} e^2 ( 25 c_{F^2 D^2 \phi^2}^{(1)} + 72 c_{F^2 D^2 \phi^2}^{(2)} ) ( s^2 + t^2 + u^2 ) \,.
\end{split}
\eeq
which gives
\begin{equation}
\label{eq: sigma photon 1}
    \begin{split}
        \Sigma^{\prime}_{yy} \cdot \Lambda^{4} =&  16 c_{F^{4}}^{(1)} + \frac{1}{4 \pi^2} c_{F^2 \phi^2}^{2} \left( - \frac{3}{2} - 2 \log \left( \frac{s_0}{\mu^2} \right) + 4 \right)  \\
        & \quad - \frac{1}{96 \pi^2} e^2 c_{F^2 D^2 \phi^2}^{(1)} \left( -\frac{3}{2} - 2 \log \left( \frac{s_0}{\mu^2} \right) - 2 \log \left( \frac{m^2_\gamma}{\mu^2} \right) + 8 \right) \\
        & \quad - \frac{ 1 }{144 \pi^2} e^2 ( 25 c_{F^2 D^2 \phi^2}^{(1)} + 72 c_{F^2 D^2 \phi^2}^{(2)} ) \,. 
    \end{split}
\end{equation}
Notably, infrared poles of the form $1 / \epsilon$ are absent from $\Sigma^{\prime}_{yy}$. A natural interpretation is that the absence of photon self-interactions prohibits real emissions from the initial and final state photons, in which case the virtual diagrams must be free of IR poles by themselves.  We have also introduced a fictitious photon mass $m_\gamma$ to regulate the IR divergence associated with the $\log(-t)$ term.   
The one-loop dim-8 contributions in \autoref{eq: sigma photon 1} from $c_{F^2 D^2 \phi^2}^{(1)}$ and $c_{F^2 D^2 \phi^2}^{(2)}$ are actually two-loop effects in the UV theory since $c_{F^2 D^2 \phi^2}^{(1)}$ and $c_{F^2 D^2 \phi^2}^{(2)}$ are themselves one-loop generated in the UV.
On the other hand, the term proportional to $c_{F^2 \phi^2}^{2}$ (the square of a one-loop generated dim-6 coefficient) is a three-loop effect in the UV theory. 
As such, the bound $c_{F^{4}}^{(1)}\geq0$ is still valid at the one-loop level in the UV theory, as we already expected.  
One may also worry that a large log term in \autoref{eq: sigma photon 1} could potential spoil the bound $c_{F^{4}}^{(2)}\geq0$.  However, as we already emphasized in \autoref{sec:phosca}, the dispersion relation holds order-by-order in the perturbative expansion, so the bound is valid as long as we consistently include all contribution up to the desired loop order in the UV theory.

For $yx\to yx$, similarly, the dim-4 one-loop contribution (also shown in \autoref{sec:aaaadim4}) can be subtracted.    
Meanwhile, the dim-6 one-loop amplitude is proportional to $t$ whose forward limit is zero:
\begin{equation}
    \mathcal{A}^{[6],\text{1-loop}}_{yx \to yx} = - \frac{1}{\Lambda^2} \frac{e^2}{2\pi^2}c_{F^2\phi^2}t \, .
\end{equation}
Therefore, $\Sigma'_{yx}$ (defined as in \autoref{eq:sigmaa1xs} with $yy$ replaced by $yx$) also contains only the dim-8 contributions.  The dim-8 one-loop amplitude is 
\begin{equation}
\label{eq:4-gamma-loopamp-2}
    \begin{split}
        \mathcal{A}^{[8],\text{1-loop}}_{yx \to yx} =& \frac{1}{\Lambda^4} \frac{1}{96\pi^2}\Big[ e^2 c_{F^2 D^2\phi^2}^{(1)} \Big( s(u-t)B_0(s)+u(s-t)B_0(u)) \Big)\\
        &-B_0(t) \Big( e^2 c_{F^2 D^2 \phi^2}^{(1)} (s^2+u^2) + 24 c_{F^2\phi^2}^2 t^2 \Big) \Big] \\ 
        &+ \frac{1}{\Lambda^4}\frac{1}{288 \pi^2} e^2 \Big(
        c_{F^2 D^2 \phi^2}^{(1)}(15 s^2 + 44 s u + 15 u^2)+ 72 c_{F^2 D^2 \phi^2}^{(2)} t^2\Big) \,,
    \end{split}
\end{equation}
which gives
\begin{equation}
\label{eq: sigma photon 2}
    \begin{split}
        \Sigma^{\prime}_{yx} \cdot  \Lambda^{4}  =&  16 c_{F^{4}}^{(2)} + \frac{1}{96\pi^2} e^2 c_{F^2 D^2\phi^2}^{(1)} \left( \frac{3}{2} + 2 \log \left( \frac{s_0}{\mu^2} \right) + 2 \log \left( \frac{m^2_\gamma}{\mu^2} \right) - 8 \right) 
        - \frac{7}{144 \pi^2} e^2 c_{F^2 D^2 \phi^2}^{(1)}  \,, 
    \end{split}
\end{equation}
where we have also introduced a fictitious photon mass $m_\gamma$.  The contribution from $c_{F^2 D^2 \phi^2}^{(1)}$ is again a two-loop effect in the UV theory, so the bound $c_{F^{4}}^{(2)} \geq 0$ is still valid up to the one-loop order in the UV theory.  As explicit checks, we also note that both $c_{F^{4}}^{(1)} \geq 0$ and $c_{F^{4}}^{(2)} \geq 0$ are satisfied by the two models in \autoref{subsec:topdow}, as shown in \autoref{sec:matching}.

\begin{figure}[t]
    \centering
    \includegraphics[width=0.7\linewidth]{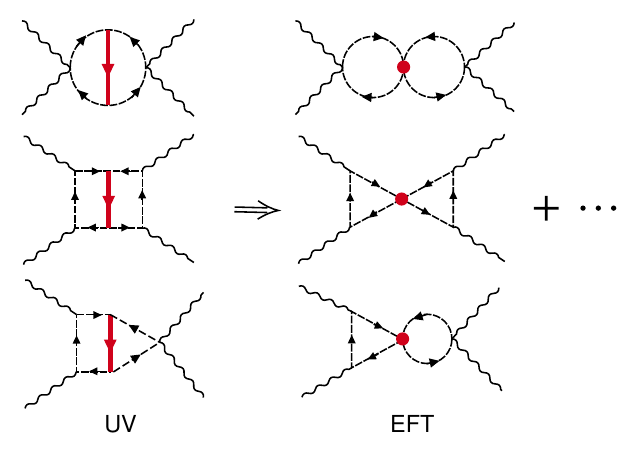}
    \caption{Diagrams for the two-loop contribution from $c_{D^4 \phi^4}^{(1)}$ to the $\aaaa$ amplitude (shown on the right side),  which is missing in the dispersion relations in \autoref{eq: sigma photon 1} and \autoref{eq: sigma photon 2}.  It can be generated with the heavy scalar exchange (represented by the thick red line) in Model~II, shown on the left side.
    Note the same diagrams in the UV also generates other tree-level and one-loop diagrams in the EFT from matching (represented by the ``+ ...''). It also contributes to $\sigma(\gamma\gamma \to \phi \phi^\dagger)$ on the RHS of the dispersion relation \autoref{eq:Sigmapos} if cut at two light scalars in the loop.
   }
    \label{fig: four gamma two loop}
\end{figure}

The dispersion relations in \autoref{eq: sigma photon 1} and \autoref{eq: sigma photon 2} provide a verification of the tree-level bounds in \autoref{eq:4gamma-treebound-1} and \autoref{eq:4gamma-treebound-2}. 
It would be interesting to see if other useful information could be extracted from them.  Since they are free from the $1/\epsilon$ IR poles (unlike \autoref{eq:phi-gamma-sigma}), naively one might imply the bounds $\beta( c_{F^{4}}^{(1)} ) \leq 0$ and $\beta( c_{F^{4}}^{(2)} ) \leq 0$ following the same logic as in \autoref{subsec:Imp}. 
However, other issues still arises as we go beyond the one-loop level in the UV theory.  First, the cross section $\sigma(\gamma\gamma \to \phi\phi^\dagger)$ would contain contributions from heavy particles, so the subtraction in \autoref{eq:sigmaa1xs} cannot be performed.\footnote{In other words, if we just subtract the dim-4 contributions, there is no guarantee that the rest of $\Sigma$ is still positive.}  Furthermore, if $c_{D^4 \phi^4}^{(1)}$ (or $c_{D^4 \phi^4}^{(2)}$) is generated at the tree-level in the UV theory, then its two-loop contribution in the EFT would be at the same order as the one-loop contributions in \autoref{eq: sigma photon 1} and \autoref{eq: sigma photon 2} (both at the two-loop order in the UV theory) and should be included as well, as illustrated in \autoref{fig: four gamma two loop}.

As explicit checks we could also compute the $\beta$-functions of $c_{F^{4}}^{(1)}$ and $c_{F^{4}}^{(2)}$.  At the one-loop order in the EFT, we have
\begin{equation}
\label{eq: beta function four photon}
\begin{split}
    \beta( c_{F^{4}}^{(1)} ) = \frac{1}{192 \pi^2}( e^2 c_{F^2 D^2 \phi^2}^{(1)} - 12 c_{F^2\phi^2}^2 )\,, \quad\quad\quad \beta( c_{F^{4}}^{(2)} ) = \frac{1}{192 \pi^2} e^2 c_{F^2 D^2\phi^2}^{(1)}\,.
\end{split}
\end{equation}
If the tree-level positivity bound $c_{F^{2}D^{2}\phi^{2}}^{(1)} < 0$ is valid, then \autoref{eq: beta function four photon} would imply that both $\beta$-functions are smaller than zero.   
This would be the case for Model~I in \autoref{subsec:topdow}, where all the higher dimensional operators are generated at the loop level, and \autoref{eq: beta function four photon} is also valid up to the two-loop level in the UV model. 
Indeed, in Model~I we have
\begin{equation}
\begin{split}
    \beta( c_{F^{4}}^{(1)} ) = -\frac{e^4 g^2}{7741440 \pi^4} <0 \,, \quad\quad\quad
    \beta( c_{F^{4}}^{(2)} ) = -\frac{e^4 g^2}{7741440\pi^4} <0 \,. 
\end{split}
\end{equation}
Interestingly, the statement above is stronger than what one could derive from the dispersion relation.  
This naturally raises the question that whether one could obtain a refined version of the dispersion relation which bounds the $\beta$-functions under these assumptions.  
On the other hand, in Model~II in \autoref{subsec:topdow} the operator $c_{D^4 \phi^4}^{(1)}$ is generated at the tree level, and the na\"ive tree-level bound $c_{F^{2}D^{2}\phi^{2}}^{(1)} < 0$ does not necessarily hold.  If we just use the EFT one-loop $\beta$-functions in \autoref{eq: beta function four photon} without including the two-loop contribution from $c_{D^4 \phi^4}^{(1)}$, we obtain
\begin{equation}
\begin{split}
    \beta( c_{F^{4}}^{(1)} ) = \frac{e^4 g^2}{2048\pi^4} > 0 \,, \quad\quad\quad
    \beta( c_{F^{4}}^{(2)} ) = \frac{e^4 g^2}{2048\pi^4}  >0 \,,
\end{split}
\end{equation}
which are a clear counterexample of the na\"ive bounds above.


\section{Conclusion}
\label{sec:con}

In this paper, we derived the one-loop positivity bounds of the $\apap$ and $\aaaa$ amplitudes in Scalar-QED EFT and studied their implications on the corresponding dim-8 operator coefficients. It is important to go beyond the tree-level positivity bounds for these amplitudes since they could only be generated at the one-loop level in the UV theory under generic assumptions (weakly coupled heavy particles with spin $\leq1$).  In contrast to the pure scalar EFT considered in Ref.~\cite{Ye:2024rzr}, gauge invariance imposes strong restrictions on the possible loop structures in Scalar-QED EFT, which has important implications for the positivity bounds.  The potential IR divergences also require careful treatments.  In particular, divergences of the form $\log(-t)$ are regulated with a small mass term when necessary, while IR poles of the form $1/\epsilon$ are shown to be absent in the dispersion relation at the one-loop level in the UV theory. 
For $\apap$, we find that, while the tree-level bound $c_{F^2 D^2 \phi^2}^{(1)} \leq 0$ does not necessarily hold, its one-loop $\beta$-function is subject to a bound that tends to restore the tree-level bound in the IR, if it is also generated at the one-loop level in the UV theory ({\it i.e.}, the RG mixing comes from tree-level generated coefficients). 
For $\aaaa$, the tree-level bounds $c_{F^{4}}^{(1)} \geq 0$ and $c_{F^{4}}^{(2)} \geq 0$ are valid up to the one-loop level in the UV theory, since $c_{F^{4}}^{(1)}$ and $c_{F^{4}}^{(2)}$ are the only contribution to $\aaaa$ after integrating out the heavy particle in the loop.

Our results on the $\beta$-functions exhibit similar patterns to the findings in Refs.~\cite{Chala:2023jyx,Chala:2023xjy,Liao:2025npz}. However, as we already emphasized, the bounds we derive are in general only valid up to the one-loop level in the UV theory, as a result of the simple one-loop structures.
In this sense, the positivity bounds on the $\beta$-functions should be considered to be an accidental feature at one loop, rather than a fundamental property of the theory.  Nevertheless, it remains an interesting question whether any bounds on the $\beta$-functions could still be valid at higher loop orders, perhaps under some additional model assumptions.

It is also desirable to generalize our framework to more realistic theories, such as QED (EFT) or SMEFT, which we leave for future studies.  The results could have important implications for experimental tests of positivity bounds, such as the ones related to $e^+e^-\to \gamma\gamma$ or $\gamma\gamma \to \ell^+\ell^-$ considered in Refs.~\cite{Gu:2020ldn,Gu:2023emi}.  On the other hand, given the difficulty in relating positivity bounds to operator coefficients at the loop level, perhaps a better approach would be to bypass the Lagrangian framework and directly link positivity bounds with on-shell amplitudes that are directly related to observables.  This goes in the same direction as the $S$-matrix bootstrap program (see {\it e.g.} Ref.~\cite{Kruczenski:2022lot} for a review and references therein), which is also highly nontrivial if one goes beyond the tree level.


\subsection*{Acknowledgments}

We thank Cyuan-Han Chang, Xu Li, Chia-Hsien Shen and Shuang-Yong Zhou for useful discussions and valuable comments on the manuscripts.  This work is supported by the National Natural Science Foundation of China (NSFC) under grant No.\,12035008 and No.\,12375091, and the Innovation Program for Quantum Science and Technology under grant No.\,2024ZD0300101. 
JG also thanks the organizers of the Positivity, Amplitudes, and Phenomenology Workshop at CERN where many stimulating discussions took place.


\appendix

\section{Scalar one-loop integrals}
\label{sec:masint}

In this appendix we list basic scalar one-loop integrals for massless theories in $D = 4 - 2\epsilon$ dimensions.
These results can be found in \cite{Bern:1993kr}, and the repository of basic scalar integrals with internal masses can be found in \cite{Ellis:2007qk}.
The prefactor $e^{\epsilon \gamma_{\text{E}}}$ is introduced to remove Euler constant $\gamma_{\text{E}}$ from the final result, because we chooses the modified minimal subtraction scheme $\overline{\text{MS}}$.
Following the notation of Passarino-Veltman master integrals in \texttt{FeynCalc}, all momenta $p_{i}$ are incoming, $p_{1} + p_{2} + p_{3} + p_{4} = 0$.  

\subsection*{Bubble integral}
The bubble integral
\begin{equation}
    \begin{split}
        B_{0}( p_{1}^{2}, m_{0}^{2}, m_{1}^{2} ) = e^{\epsilon \gamma_{\text{E}}} \int \frac{ d^{D} \ell }{ i \pi^{D/2} } \frac{ 1 }{ \ell^{2} - m_{0}^{2} + i \varepsilon } \frac{ 1 }{ ( \ell + p_{1} )^{2} - m_{1}^{2} + i \varepsilon } \,
    \end{split}
\end{equation}
is given by (in the massless case)
\begin{equation}
    B_{0}( p_{1}^{2}, 0, 0 ) = \frac{1}{\epsilon} - \log\left( \frac{ -p_{1}^{2} -i\varepsilon }{ \mu^{2} } \right) + 2 + \mathcal{O}(\epsilon) \,.
\end{equation}
It has UV divergences, but no IR divergences.
We take the shorthand $B_0(s) = B_0(s, 0, 0)$.

\subsection*{Triangle integral}
The triangle integral
\begin{equation}
    \begin{split}
        &C_{0}( p_{1}^{2}, p_{2}^{2}, (p_{1} + p_{2})^{2}, m_{0}^{2}, m_{1}^{2}, m_{2}^{2} ) \\
        =& e^{\epsilon \gamma_{\text{E}}} \int \frac{ d^{D} \ell }{ i \pi^{D/2} } \frac{ 1 }{ \ell^{2} - m_{0}^{2} + i \varepsilon } \frac{ 1 }{ ( \ell + p_{1} )^{2} - m_{1}^{2} + i \varepsilon } \frac{ 1 }{ ( \ell + p_{1} + p_{2} )^{2} - m_{2}^{2} + i \varepsilon } \,
    \end{split}
\end{equation}
is given by
\begin{equation}
    \begin{split}
        &C_{0}( 0, 0, (p_{1} + p_{2})^{2}, 0, 0, 0 ) \\ 
        =& \frac{1}{s} \left( \frac{1}{\epsilon^2} - \frac{1}{\epsilon} \log \left( \frac{ - s -i\varepsilon }{ \mu^{2} } \right) + \frac{1}{2} \log^{2} \left( \frac{ - s -i\varepsilon }{ \mu^{2} } \right) - \frac{1}{2} \zeta_{2} \right) + \mathcal{O}(\epsilon) \,, 
    \end{split}
\end{equation}
where $\zeta_2 = \frac{\pi^2}{6}$.
Power counting of the superficial degree of divergence shows that the triangle integral cannot have UV divergences.
The $1 / \epsilon$ and $1 / \epsilon^2$ terms come from the IR (soft and collinear) divergences. 
We take the shorthand $C_0(s) = C_0(0, 0, s, 0, 0, 0)$.

\subsection*{Box integral}
The box integral
\begin{equation}
    \begin{split}
        &D_{0}( p_{1}^{2}, p_{2}^{2}, p_{3}^{2}, p_{4}^{2}, (p_{1} + p_{2})^{2}, (p_{2} + p_{3})^{2}, m_{0}^{2}, m_{1}^{2}, m_{2}^{2}, m_{3}^{2} ) \\
        =& e^{\epsilon \gamma_{\text{E}}} \int \frac{ d^{D} \ell }{ i \pi^{D/2} } \Big( \frac{ 1 }{ \ell^{2} - m_{0}^{2} + i \varepsilon } \frac{ 1 }{ ( \ell + p_{1} )^{2} - m_{1}^{2} + i \varepsilon } \frac{ 1 }{ ( \ell + p_{1} + p_{2} )^{2} - m_{2}^{2} + i \varepsilon } \\
        &\quad \quad \quad \quad \quad \times \frac{ 1 }{ ( \ell + p_{1} + p_{2} + p_{3} )^{2} - m_{3}^{2} + i \varepsilon } \Big) \,
    \end{split}
\end{equation}
is given by
\begin{equation}
    \begin{split}
        &D_{0}( 0, 0, 0, 0, (p_{1} + p_{2})^{2}, (p_{2} + p_{3})^{2}, 0, 0, 0, 0 ) \\
        =& \frac{ 1 }{ s u } \Bigg( \frac{ 4 }{ \epsilon^2 } - \frac{ 2 }{ \epsilon } \log \left( \frac{ - s -i\varepsilon }{ \mu^{2} } \right) - \frac{ 2 }{ \epsilon } \log \left( \frac{ - u -i\varepsilon }{ \mu^{2} } \right) \\
        & \quad + \log^{2} \left( \frac{ - s -i\varepsilon }{ \mu^{2} } \right) + \log^{2} \left( \frac{ - u -i\varepsilon }{ \mu^{2} } \right) - \log^{2} \left( \frac{ - s -i\varepsilon }{ - u -i\varepsilon } \right) - 8 \zeta_2 \Bigg) + \mathcal{O}(\epsilon) \, .
    \end{split}
\end{equation}
For the same reason, the box integral has only IR divergences.
We take the shorthand $D_0(s,u) = D_0(0, 0, 0, 0, s, u, 0, 0, 0, 0)$.


\section{Full one-loop matching results}
\label{sec:matching}

In this appendix we provide the one-loop matching results for the two models in \autoref{subsec:topdow}, obtained with the \texttt{Matchete} package~\cite{Fuentes-Martin:2022jrf}.    Only the coefficients that contributions to $\apap$ and $\aaaa$ up to the one-loop level in the EFT (those listed in \autoref{eq:c1loop}) are included here.  For generality, we have kept the renormalization scale $\mu$ in the expressions, which should be set to the matching scale $\mu = M$ when matching the UV model to the EFT.  The coefficients could then be run down to a lower scale with the RGEs in the EFT.  Note that, if we just keep the leading term without resumming the log terms, the results from matching and running would be exactly the ones in \autoref{eq:match1-2} and \autoref{eq:match2-2} below.  
We have also set $\Lambda=M$.

\subsubsection*{Model~I} with the Lagrangian in \autoref{eq:testUV1}:  
\begin{equation}
\label{eq:match1-2}
    \begin{split}
        &c_{F^{2}D^{2}\phi^{2}}^{(1)} = - \frac{1}{16\pi^2} \frac{1}{2520} g^2 e^2 \,, \quad c_{F^{2}D^{2}\phi^{2}}^{(2)} = \frac{1}{16\pi^2} \frac{1}{10080} e^2 ( 23 g^2 - 56 \lambda_3 ) \,, \\
        &c_{F^{4}}^{(1)} = \frac{1}{16\pi^2} \frac{7}{23040} e^4 \,, \quad c_{F^{4}}^{(2)} = \frac{1}{16\pi^2} \frac{1}{23040} e^4 \,, \\
        &c_{D^4 \phi^4}^{(1)} = \frac{1}{16 \pi^2} \frac{1}{5040} (6e^4 - 8 g^2 e^2 + 3 g^4) \,, \\ 
        &c_{D^4 \phi^4}^{(2)} = \frac{1}{16\pi^2} \frac{1}{2520}(-3 e^4 + 4 g^2 e^2 + 9 g^4 -56 g^2 \lambda_3 + 84 \lambda_3^2)\,, \\
        &c_{F^2 \phi^2} = \frac{1}{16 \pi^2} \frac{1}{480} e^2 \left( -3 g^2 + 10 \lambda_3 \right)\,.
    \end{split}
\end{equation}
Note that all operators are generated at one-loop level in this model, so the coefficients have no running at the one-loop level in the UV model.

\subsubsection*{Model~II} with the Lagrangian in \autoref{eq:testUV2}:
\begin{equation}
\label{eq:match2-2}
    \begin{split}
        &c_{F^{2}D^{2}\phi^{2}}^{(1)} = \frac{1}{16\pi^2} \frac{1}{6} g^2 e^2 \left(9 + 4 \log ( \mu^2 / M^2 ) \right)\,, \\
        &c_{F^{2}D^{2}\phi^{2}}^{(2)} = - \frac{1}{16\pi^2} \frac{1}{360} e^2 \left[ 32 \lambda_3 + 3 g^2 \left( 9 + 20 \log ( \mu^2 / M^2 ) \right) \right] \,, \\
        &c_{F^{4}}^{(1)} = \frac{1}{16\pi^2} \frac{7}{90} e^4, \quad c_{F^{4}}^{(2)} = \frac{1}{16\pi^2} \frac{1}{90} e^4 \,, \\
        &c_{D^4 \phi^4}^{(1)} = g^2 + \frac{1}{16 \pi^2} \frac{1}{630} \Bigg\{ 12 e^4 - 21 g^2 e^2 \left( 189 + 220 \log ( \mu^2 / M^2 ) \right)  \\
        & \quad \quad \quad - 35 g^2 \Big[ - 18 ( 3 \lambda_2 - 4 \lambda_3 ) \left( 1 + \log ( \mu^2 / M^2 ) \right) + g^2 \left( 29 + 3 \log ( \mu^2 / M^2 ) \right) \Big] \Bigg\} \,, \\ 
        &c_{D^4 \phi^4}^{(2)} = \frac{1}{16\pi^2} \frac{1}{630} \Bigg\{ -12 e^4 + 21 \lambda_3^2 - 7 g^4 \left( 277 + 150 \log ( \mu^2 / M^2 ) \right) \\
        & \quad \quad \quad - 7 g^2 e^2 \left(  473 + 300 \log ( \mu^2 / M^2 ) \right) + 28 g^2 \left( - \lambda_3 + 5 \lambda_1 \left( 11 + 6 \log ( \mu^2 / M^2 ) \right) \right) \Bigg\} \,, \\
        &c_{F^2 \phi^2} = \frac{1}{16 \pi^2} \frac{1}{3} e^2 \left( - g^2 + \lambda_3 \right) \,.
    \end{split}
\end{equation}


\section{Dim-4 one-loop $\aaaa$ amplitudes}
\label{sec:aaaadim4}

The dim-4 one-loop amplitude of $\gamma^{y}\gamma^{y} \to \gamma^{y}\gamma^{y}$, generated by a scalar loop, is given by
\begin{equation}
\label{eq:A4yyyy}
    \begin{split}
        \mathcal{A}^{[4],\text{1-loop}}_{yy \to yy} = & \frac{ e^4 }{ 8 \pi^2 } \Big[ (s^2 + t^2 + u^2) \left( \frac{1}{su} B_0(t) + \frac{1}{tu} B_0(s) + \frac{1}{st} B_0(u) \right) \\
        &+ \frac{ s^3 ( 3 t^2 + 3 u^2 - s^2 ) }{ t^2 u^2 } C_0(s) + \frac{ t^3 ( 3 s^2 + 3 u^2 - t^2 ) }{ s^2 u^2 } C_0(t) \\
        &+ \frac{ u^3 ( 3 s^2 + 3 t^2 - u^2 ) }{ s^2 t^2 } C_0(u) \\
        &+ \frac{s^2 u^2}{t^2} D_0(s, u) + \frac{t^2 u^2}{s^2} D_0(t, u) + \frac{s^2 t^2}{u^2} D_0(s, t) \Big] - \frac{ e^4 }{ 4 \pi^2 } \,.
    \end{split}
\end{equation}
It is straightforward to verify that the UV divergences in this amplitude cancel precisely, as required by the absence of a corresponding four-photon counterterm in \autoref{eq: scalar qed d<=4}. The IR divergences that appear in \(C_0\) and \(D_0\) cancel each other completely. 
However, simple poles of the form \(1/t\) and \(1/t^2\) still appear, seemingly contradicting the result in \autoref{sec:sQED}, which asserts the absence of a \(t\)-channel simple pole. 
On the other hand, the 2-to-2 tree-level cross section
\begin{equation}
    \sigma(\gamma^{y} \gamma^{y} \to \phi \phi^{\dagger}) = \frac{ e^{4} }{ 8 \pi s } \, 
\end{equation}
is finite, which implies that $ \mathcal{A}^{[4],\text{1-loop}}_{yy \to yy}$ is well defined in the forward limit, according to the dispersion relation \autoref{eq:Sigmapos}.   
To obtain a finite and consistent result, it is essential to keep $t$ as a small quantity and perform a \textit{Taylor expansion} for the $\log(-u) = \log(s) + \log( 1 + \frac{t}{s} )$ and $\frac{1}{u} = - \frac{1}{s} \frac{1}{ 1 + \frac{t}{s} }$ terms in \autoref{eq:A4yyyy}. By doing so, the $1/t$ and $1/t^2$ terms cancels and we obtain
\begin{equation}
    \begin{split}
        A^{[4],\text{1-loop}}_{yy \to yy} \simeq&\frac{ e^4 }{ 8 \pi^2 } \Big[ - \log(-s) - \log(s) + 2 \log(-t) - 3 + \mathcal{O}\left( \frac{t}{s} \right) \Big] \, ,
    \end{split}
\end{equation}
which gives
\begin{equation}
    \label{eq:sigma yyyy dim-4}
    \oint_{\gamma} \frac{ds}{ 2\pi i } \frac{ s^{3} A^{[4],\text{1-loop}}_{y y \to y y} }{ (s^{2} + s_{0}^{2})^{3} } = \frac{2}{\pi} \int_0^{\infty} ds\,\frac{ s^4 \sigma(\gamma^{y} \gamma^{y} \to \phi \phi^{\dagger}) }{\left(s^2+s_0^2\right)^3}  = \frac{ e^{4} }{ 16 \pi^2 s_{0}^{2} } \,,
\end{equation}
consistent with the dispersion relation \autoref{eq:Sigmapos}.  This contribution is subtracted in \autoref{eq:sigmaa1xs}.

Similarly, for $\gamma^{y}\gamma^{x} \to \gamma^{y}\gamma^{x}$, the dim-4 one-loop amplitude is
\begin{equation}
\label{eq:4-gamma-1loop-d4}
    \begin{split}
        \mathcal{A}^{[4],\text{1-loop}}_{y x \to y x} = & \frac{ e^4 }{ 8 \pi^2 } \frac{1}{s^2 t^2 u^2} \Big[ 2 s^3 u t (s+2u) B_0(s) + 2 u^3 s t (2s + u) B_0(u) \\
        & + t^2 su (s^2 + t^2 + u^2) B_0(t) + t^5(3 s^2 + 3 u^2 - t^2) C_0(t) \\
        & + (s^2 + t^2 + u^2) \Big( s^4 (u-t) C_0(s) + u^4 ( s - t ) C_0(u) \Big) \\
        &- s^4 u^4 D_0(s, u) + s^4 t^4 D_0(s, t) + t^4 u^4 D_0(t, u) \Big] + \frac{e^4}{4\pi^2} \,.
    \end{split}
\end{equation}
By doing a Taylor expansion around $t=0$, it becomes
\begin{equation}
    \begin{split}
        A^{[4],\text{1-loop}}_{y x \to y x} \simeq \frac{ e^4 }{ 8 \pi^2 } \Big[ - \log(-s) - \log(s) + 2 \log(-t) + 3 + \mathcal{O}\left( \frac{t}{s} \right) \Big] \, ,
    \end{split}
\end{equation}
which is consistent with the 2-to-2 tree-level cross section
\begin{equation}
\label{eq:crosec-yxphph}
    \sigma(\gamma^{y} \gamma^{x} \to \phi \phi^{\dagger}) = \frac{ e^{4} }{ 8 \pi s } \, ,
\end{equation}
and its contribution to the dispersion relation can also be subtracted.


\bibliographystyle{JHEP}

\bibliography{posloop}

\end{document}